\documentclass{elsart}
\usepackage{amssymb}
\usepackage{amsmath}
\usepackage{graphics}

\usepackage{epsfig}
\usepackage{amssymb}

\def\bfg #1{{\mbox{\boldmath $#1$}}}
\begin{document}
\begin{frontmatter}
\begin{center}
{\Large \bf Parity and spin of the $\Theta^+$ pentaquark 
 in the $NN\to  Y\Theta^+$ reaction
 at the threshold
}
\end{center}

\begin{center}
{Yu.N.~Uzikov}
\footnote{E-mail address: uzikov@nusun.jinr.ru
}

{\it Joint Institute for Nuclear Research, LNP, 141980 Dubna,
 Moscow Region, Russia}
\end{center}

\begin{abstract}
{ Spin observables  of a binary reaction $1+2\to 3+4$
 are discussed at the threshold in  general form
 using P-parity  and angular momentum conservation. 
 General formulae for  polarization transfer,
 spin-spin correlation parameters in the initial  and final
 states   and induced tensor polarization
 are derived for arbitrary spins of  participating particles.
 This formalism is worked out in detail for the 
 $NN\to Y\Theta^+$ reaction.  Assuming that  the spin of the
 pentaquark $\Theta^+$ takes the values
 $\frac{1}{2},\,\frac{3}{2}$ and  $\frac{5}{2}$,  whereas 
 the spin of the hyperon $Y$  equals  $\frac{1}{2}$, explicite 
 formulae are obtained for the observables  in terms of few
 non-vanishing at the threshold spin amplitudes separately
 for the spin-singlet and spin-triplet initial NN states.
 In case of all  particles in the $NN\to Y\Theta^+$ reaction
 have  the spin-$\frac{1}{2}$  a full spin-structure
 for totally polarized cross sections is derived.  
 Some of the obtained spin observables strongly depend on the
 intrinsic P-parity of the $\Theta^+$ and the total isospin
 of the reaction. Therefore, measurements of  these observables
 allow one    to determine  the P-parity of the 
 pentaquark $\Theta^+$  in a model independent way
 for any spin of the $\Theta^+$.
}

\end{abstract}  

\begin{keyword}


\begin{PACS}
\end{PACS}
\end{keyword}

\end{frontmatter}
\baselineskip 4ex
\newpage

\section{Introduction}

  Experimental indications
 \cite
 {Nakano,Barmin,Stepanyan,Kubarovsky,Elsa,hermes,SVD,COSYTOF,aslanyan,troyan}
 to existence of the  exotic baryon $\Theta^+(1540)$
 with the strangeness $S=+1$
 stimulated an intensive theoretical discussion during the past year.
 A minimal number of constituent quarks 
 in the $\Theta^+(1540)$, compatible with its positive strangeness and the
 baryon charge, is five  and corresponds to the quark content  $uudd{\bar s}$.
 The structure of the $\Theta^+(1540)$ and its quantum
 numbers originally predicted in the chiral
 soliton model \cite{dpp} ($j^\pi_\Theta=\frac{1}{2}^+$),
 have been considered yet in several
  quark models with a different type of
 quark-quark interaction, as well as within the  QCD sum rules and
 Lattice QCD. For recent review see, for example, Refs. \cite{review}.
 A surprisingly   narrow width 
 of the $\Theta^+$ ($\Gamma_\Theta \approx 15$ MeV) 
 predicted by  the chiral soliton model \cite{dpp}, 
 now  is expected to be about of
 1 MeV or less  according to the recent
 analysis \cite{width}
 of  the $KN$- and $Kd$- scattering data and the DIANA
 data \cite{Barmin} on the $\Theta^+$ production \cite{sibirtsev}.
 This
 looks also as an exotic itself in view of the open decay channel 
 $\Theta^+\to KN$  with the same quark content in the $KN$ system as in the
 $\Theta^+$.
 The width of the $\Theta^+$ is now a challenge for any theoretical
 approach 
\cite{karlinerlipkin,kochelev,hosaka32,jaffejain,ioffeoganesian,nishikawa},
including the chiral soliton model also \cite{ellis,diakonov}. 
 The intrinsic parity of the $\Theta^+$ is of
 particular importance to understand the nature of this baryon. 
 A positive  parity of the
 $\Theta^+$ is predicted by the chiral soliton model \cite{dpp}. On
 the contrary, naive quark shell models give the
 negative parity for the S-wave  ground state of the quark system
 $uudd\bar{s}$. Quark-quark 
 interaction could lead to a P-wave state and, hence, to positive parity 
 of the $\Theta^+$ \cite{jaffewilczek,karlip,lattice2}.
  In  Lattice QCD,
  both the positive  \cite{lattice2}  and  negative \cite{lattice1} 
  parity  of the $\Theta^+$ are obtained
  although in Refs. \cite{lattice0} the $\Theta^+(1540)$
  resonance state is not found.  
  A possible correlation between the  positive parity of the $\Theta^+$ and
 its narrow width is discussed in  \cite{carlson}  within a constituent quark
 model. 
 Experimental  determination of 
 the quantum numbers of the $\Theta^+$ will be essential for establishing of
 the underlying dynamics of the $\Theta^+$.
\footnote{  Besides of the $\Theta^+(1540)$ observation,
 an indication to existence 
 of two others explicitely exotic baryons were also reported: 
 $\Xi^{--}(1860)$ \cite{na49} and $\Theta_c^0(3099)$ \cite{h1}. However, in
 several experiments performed mainly at high energies the states
$\Theta^+(1540)$, $\Xi^{--}(1860)$,  $\Theta_c^0(3099)$ were not found.
 For review of positive and null experimental results one can see Refs. 
\cite{pochodzalla,hicks}. At present, null results obtained at high energies
 do not mean that  
the experiments 
\cite
 {Nakano,Barmin,Stepanyan,Kubarovsky,Elsa,hermes,SVD,COSYTOF,aslanyan,troyan}
 are incorrect. Most likely,  the $\Theta^+(1540)$ production mechanism
 depends on energy and transferred  momentum  in such a way that at high
 energies this state is non-visible \cite{titov}. New experiments with high
 statistics are planned \cite{hicks} to get definite answer about existence of
 the exotic baryons.}

 Several methods, based on certain assumptions on the production mechanism,
 were suggested for determination of the P-parity
 of the $\Theta^+$, $\pi_\Theta$,
 in  Kp-scattering \cite{oset} and
 photoproduction reactions \cite{photoproduct}.
 Model independent methods were also considered \cite{rekalo1}. 
 We discuss here  the $\Theta^+$ production  in $NN$-collision.   
 Model independent methods for determination of the P-parity of the
  $\Theta^+$ in the reaction $NN\to Y\,\Theta^+$ were
 suggested in Refs. \cite{thomas,hanhart,uzikov,rekalo2}. These
 methods are based on such   general properties of the reaction 
  amplitude  as  angular momentum and P-parity conservation and 
  on the generalized Pauli principle for  nucleons.
  Assuming  for the spin of the $\Theta^+$ the value 
 $j_\Theta =\frac{1}{2}$, 
 it was shown in Ref.\cite{hanhart} that
 the sign of the spin-spin correlation parameter $C_{y,y}$ 
 unambiguously determines the P-parity of the $\Theta^+$ 
 in the reaction $pp\to\Sigma^+\Theta^+$. 
 Another definite
 correlation between  $C_{y,y}$ and $\pi_\Theta$
 holds  for the $pn\to \Lambda^0 \Theta^+$ reaction 
 \cite{uzikov,rekalo2} if the isospin of the $\Theta^+$ 
 equals  zero.
 Furthermore, a
 measurement of the spin transfer coefficients $K_y^y=K_x^x$ or
 $K_z^z$ in  these reactions also allows to
 determine the P-parity unambiguously  \cite{uzikov,rekalo2}.
 Obviously, a  double-spin measurement near the threshold
 is  a challenge for experiment. In this connection, one should  note
 that   measurements of the polarization transfer from the initial
 nucleon to the
 hyperon  in the reaction $NN\to Y\Theta^+$ can be performed 
 by  single spin experiments with only
 polarized beam or target, since the polarization of the hyperon
 can be measured via its weak decay.
 
  The results of Refs. \cite{thomas,hanhart,rekalo2}
 are based on the assumption that the spin of the $\Theta^+$ 
 is equal to $\frac{1}{2}$. Up to now the experimental value of the
  spin of the $\Theta^+$
 is  not known, as well as its P-parity.
 Analysis  of the angular distribution
 in the system $nK^+$ in the final state of the reaction $np\to npK^+K^-$
 measured in \cite{troyan},
 leads the authors of Ref.\cite{troyan} to a conclusion
 that the minimal  value of the  spin of the $\Theta^+(1540)$ is
 $\frac{5}{2}$. Furthermore,
 within some quark models the pentaquark being the spin-$\frac{1}{2}$ baryon
 has partners with the spin  $\frac{3}{2}$. According to
 Ref. \cite{closedudek}, the spin-$\frac{3}{2}$ partner can have  a mass
 very close to the $\Theta^+$. As was noted recently in
 Ref. \cite{hosaka32},  the $(0s)^5$ configuration of the $uudd{\bar s}$
 pentaquark being in the
 $\frac{3}{2}^-$ state  cannot decay into a KN d-wave state. Hence the
 $\frac{3}{2}^-$ state is a candidate for the observed narrow 
 $\Theta(1540)$ resonance. A possibility  that $\Theta^+$ is the
 $\frac{3}{2}$-spin state was discussed also in Refs. 
 \cite{jaffejain,nishikawa,takeuchi,huang}.
 At last, higher spins are allowed for
 the pentaquarks $\Theta^*$ in higher SU(3) multiplets. 

 Therefore a generalization of the  approach \cite{thomas,hanhart}  
 to arbitrary spin of the $\Theta^+$ is an important task.
 This problem was discussed in Ref. \cite{uzikov2}.  
 For arbitrary spin 
 of the $\Theta^+$,
 it was shown in \cite{uzikov2} that  the signals
 for determination of  the P-parity of the $\Theta^+$ via  
 measurements of the $C_{i,j}$ and $K_i^j$
 in the reaction $NN\to Y\Theta^+$ are the same 
 as for the case of the minimal spin  $j_\Theta=\frac{1}{2}$ 
 \footnote{Independently    
 this problem  was considered in Ref. \cite{rekalo3}
 in the $\bfg \sigma$-
 representation for the transition amplitude.}.

 In this work  we develop  the method of Refs. \cite{uzikov,uzikov2}
 for calculation of $C_{i,j}$ and $K_i^j$
 for  arbitrary spins of
 {\it all} participating particles in the binary reaction $1+2\to 3+4$.
 Furthermore,  we obtain formulae for the spin-spin correlation parameters
 in the final state of the binary reaction. In this case  beam and target
 are assumed to be unpolarized. For the case of  $j_{\Theta} >\frac{1}{2}$
 we consider also 
 the tensor
 polarization of the $\Theta^+$, $t_{J0}$,
 both for polarized and unpolarized beam.
 For even rank $J$,  tensor polarization  
 can be measured via analysis of angular distribution in the strong
 decay $\Theta^+\to N+K$.
 We show that  $t_{J0}$ induced by polarized beam
 depends on the P-parity of the $\Theta^+$.
 The derived general formulae can be applied for
 {\it any binary reaction }near threshold.
 For  the $NN\to Y\Theta^+$ reaction  we present detailed  formulae  
 assuming  $j_\Theta=\frac{1}{2},\, \frac{3}{2}$ and $\frac{5}{2}$.
 In case of all participants  in the
 $NN\to Y\Theta^+$ reaction  are  the spin-$\frac{1}{2}$ particles
 we derive  totally polarized cross sections  for the  isospin
 $T=0$ and $T=1$ of the $NN$ channel, taking into account  all
 polarizations  in the initial and final states.
 Analysis is based on common properties of the reaction amplitude and 
 the standard  technique of the spin-tensor operators \cite{bilenky}.
 For the case of all particles are the spin-$\frac{1}{2}$ baryons
 we apply also the $\bfg \sigma$- representation to compare the results
 with a general method.
 In addition, we develop a non-standard method,
 which allows us  to derive an elegant formula 
 for $C_{i,j}$ for the 
 binary reaction with  initial particles being of  spin-$\frac{1}{2}$
 and final particles of arbitrary spins.

 The paper is organized in the following way. In  section  2  the 
 partial wave expansion is given for the transition amplitude 
 near the threshold.
 The spin-spin correlation parameters in the initial state
 of the reaction  $NN\to Y\,\Theta^+$ are discussed in  section 3 using
 a non-standard method. In section 4  a general method is elaborated
 to derive formulae for the spin-transfer
 coefficients,  tensor polarization induced by polarized or
 non-polarized beam,   and spin-spin correlation coefficients
for arbitrary spins 
 in the binary  reaction  $1+2\to 3+4$.   
  Explicite formulae are given for different values of 
 the $\Theta^+$ spin. 
 In  section 5
 we  derive  the  full spin structure of the cross section of 
 the reaction $NN\to Y\Theta^+$  for the case  of the  spin-$\frac{1}{2}$
 particles.
 The  results are discussed in  section 6. 
 Some technical details are explained in Appendix. 

\section{Transition amplitude}
\label{sect1}

 Assuming  dominance of the $s$-wave in the
 relative motion in the final system,
 the most general expression for the amplitude of the binary reaction
 $1+2 \to 3+4$ at the threshold 
 can be written as \cite{GW}
\begin{eqnarray}
\label{tfi}
T_{\mu_1\,\mu_2}^{\mu_3\,\mu_4}=\sum_{J\,M \,\atop
 S\,M_S\,L\,m}
(j_1\mu_1\,j_2 \mu_2|SM_S)
(j_3\mu_3\, j_4\, \mu_4|J\,M)\times \nonumber \\
\times (S\,M_S\,L\,m|J\,M)
Y_{Lm}({\hat {\bf k}}) a_J^{LS}.
\end{eqnarray}
 Here $j_i$ and $\mu_i$ are the spin of the {\it i}-th particle and its  
 z-projection, $J$ and $M$ are the total angular
 momentum and its z-projection; 
 $S$ and $L$  are the  spin  and orbital momentum of the
  initial system, respectively, and $M_S$ and $m$ are the
 corresponding z-projections.
 The coupling scheme for angular momenta is given by
\begin{eqnarray}
\label{couplspin}
 {\hat {\bf S}}={\hat {\bf j}}(1)+{\hat {\bf j}}(2), \, \, 
 {\hat {\bf S}}+{\hat {\bf L}}={\hat {\bf J}}, \, \,
 {\hat {\bf j}}(3)+{\hat {\bf j}}(4)= {\hat {\bf J}},
\end{eqnarray}
 where ${\hat {\bf j}}(i)$ is the spin-operator of 
 the {\it i}-th particle.
 Information on the reaction dynamics is
 contained in  the complex  amplitudes 
 $a_J^{L\,S}$. The sum over $J$ in Eq. (\ref{tfi}) is restricted
 by the conditions $J=j_3+j_4, j_3+j_4-1,...,|j_3-j_4|$.
 We choose the z-axis along the unit  vector of the initial momentum
 $  {\hat {\bf  k}}$, therefore 
$Y_{Lm}({\hat {\bf k}})= \sqrt{(2\,L+1)/4\pi}\,\delta_{m\,0}$.
 Due to P-parity conservation,  the orbital
 momentum $L$ in Eq. (\ref{tfi}) is restricted by the condition
 $(-1)^L=\pi$, where $\pi=\pi_1\,\pi_2\,\pi_3\pi_4$, and $\pi_i$
is the intrinsic P-parity  of the {\it i}-th particle.
 We consider here mainly  
 transitions without mixing the total isospin $T$ in this reaction
{\footnote {The isospin mixing is possible, for example, in the
 reaction $p+n\to \Sigma^0+\Theta^+$, if  the $\Theta^+$
is an isotriplet. In this case the P-parity cannot be determined
by using the method in question.}.
Using the   generalized  Pauli principle, one can find that 
for given  values of $T$ and $\pi$
 the spin of the initial nucleons $S$ is fixed 
 unambiguously by the following relation
\begin{equation}
\label{pauli}
 (-1)^S=\pi (-1)^{T+1}.
\end{equation} 
 Therefore, in order to determine the
 P-parity $\pi$ of the system at  a given isospin $T$, it is sufficient 
 to determine  the spin of the NN-system 
 in the  initial state of this reaction.

  Let us to determine the number of the spin amplitudes $a_J^{LS}$ for
  a particular case of $j_3=\frac{1}{2}$ and $j_4$ being half-integer,
  $j_4=\frac{1}{2},\,\frac{3}{2},\frac{5}{2}, \dots$.
For this case there are two total angular momenta
$J_p=j_4+\frac{1}{2}$ and $ J_m=j_4-\frac{1}{2}$. For the spin-singlet
initial state $S=0$  only one orbital momentum  is allowed, $L=J$, and 
therefore there is only one scalar amplitude, $a_J^{L\,S}=a_J^{J\,0}$,
where $(-1)^L=\pi$.
For $S=1$ and $j_4\geq \frac{3}{2}$ there are  three
 scalar amplitudes $a_J^{L\,1}\equiv a_J^{L}$:\\
(i) $a_{J_p}^{J_p},\, a_{J_m}^{J_m+1}$ and $  a_{J_m}^{J_m-1}$, if 
  $(-1)^{J_p}=\pi$,\\
or\\
 (ii) 
$a_{J_m}^{J_m}\, (J_m\not =0)$,  $a_{J_p}^{J_p+1}$ and $ a_{J_p}^{J_p-1}$, if 
  $(-1)^{J_p}=-\pi$.\\
 For particular case of $j_4=j_3=\frac{1}{2}$, one has $J_m=0$ and
 $J_p=1$. For this case only  two triplet amplitudes  are allowed for
 $\pi=+1$,
 i.e. $a_1^0$ and $a_1^2$, whereas  the amplitude $a_0^0$ is forbidden
 by conservation of the total angular momentum. For $\pi=-1$ 
 one has also only two spin-triplet amplitudes, one
 of them corresponds to $J=1$, $a_1^1$, and another one is allowed for 
 $J=0$, i.e. $a_0^1$.

\section{ Non-standard method}

 For $j_1=j_2=\frac{1}{2}$ and for given spin $S$,   
 one can find from Eq. (\ref{tfi}) the polarized cross section 
 $d\sigma({\bf p}_1,{\bf p}_2)$  as 
 \begin{eqnarray}
\label{polarsec}
d\sigma({\bf p}_1,{\bf p}_2)=\Phi \sum_{\mu_3\,\mu_4}
|T_{\mu_1\,\mu_2}^{\mu_3\,\mu_4}|^2= \nonumber \\
= \frac{1}{4\pi} \sum_{M}
(\frac{1}{2}\mu_1\frac{1}{2}\mu_2|SM)^2 
  \sum_{J\,M\,L\,L'}
\sqrt{(2L+1)(2L'+1)}(S\,M\,L\,0|J\,M)\times \nonumber \\
 \times (S\,M\,L'\,0|J\,M)
 \, a_J^{LS}\, (a_J^{L'S})^*,
\end{eqnarray} 
where $\Phi$ is a kinematical factor. 
 Using the relations $(\frac{1}{2}\mu_1\frac{1}{2}\mu_2|00)= 
 \chi_{\mu_1}^{+} \frac {i\sigma_y}{\sqrt{2}}\chi_{\mu_2}^{(T)+}$ and    
$(\frac{1}{2}\mu_1\frac{1}{2}\mu_2|1\lambda)=
 \chi_{\mu_1}^+\sigma_\lambda \frac {i\sigma_y}{\sqrt{2}}\chi_{\mu_2}^{(T)+}$,
 where $\sigma_i\, (i=y,\lambda$) is the Pauli matrix and $\chi_\mu$ is the
 2-spinor, one can find
\begin{equation}
\label{s0}
( \frac{1}{2}\mu_1\frac{1}{2}\mu_2|00  )^2=
\frac{1}{4}(1-{\bf p}_1\cdot {\bf p}_2), 
 \end{equation}
\begin{equation}
\label{s1}
 ( \frac{1}{2}\mu_1\frac{1}{2}\mu_2|1M )^2=
\begin{cases}
\frac{1}{4}(1+{\bf
    p}_1\cdot {\bf p}_2-2p_{1z}p_{2z}),  {\text{  $M=0$,}} \\
\frac{1}{4}[1 \pm (p_{1z}+p_{2z})
    +p_{1z}p_{2z}], \ {\text {$M=\pm 1$}}.
\end{cases} 
\end{equation}
In Eqs. (\ref{polarsec}), (\ref{s0}) and (\ref{s1}) 
   ${\bf p}_i$ is the polarization
 vector of the i-th particle with the spin $j_i=\frac{1}{2}$
 being in the pure spin state $\chi_{\mu_i}$. 
\footnote{ On the other side, Eqs. (\ref{s0}) and (\ref{s1}) can be
 considered as  the matrix elements 
 of  the projection operator $|SM><SM|$ \cite{ramachandran1}
 taken between the states
 $|\frac{1}{2}\mu_1,\frac{1}{2}\mu_2>$.}
 The unpolarized cross section is given as
\begin{equation}
\label{unpols}
d\sigma_0= \Phi\,\frac{1}{4} \sum_{\mu_1\,\mu_2 \,\mu_3\,\mu_4}
|T_{\mu_1\,\mu_2}^{\mu_3\,\mu_4}|^2= 
 \frac{1}{16\,\pi} \Phi \,\sum_{J,L} (2J+1)|a_J^{L\,S}|^2.
\end{equation}

 \subsection  {The spin-singlet initial state.}   
\label{sect31}
 Using Eqs. (\ref{polarsec}), (\ref{s0}) and (\ref{unpols})
 one can find for the spin-singlet polarized cross section
 the following formula
 \begin{equation}
 \label{s0sec}
 d\sigma({\bf p}_1,{\bf p}_2)=d\sigma_0 (1-{\bf p}_1\cdot {\bf p}_2). 
 \end{equation}
 As seen from this formula, the spin-singlet cross section is equal to
 zero, if the polarization vectors of colliding particles are parallel 
 (${\bf p}_1 \uparrow \uparrow{\bf p}_2$)
 and have maximal values
 ($|{\bf p}_1|=|{\bf p}_2|=1$).
   In notations of Ref. \cite{ohlsen}, non-zero spin-spin correlation
 parameters for this case are the following:
\begin{eqnarray}
\label{spin0cij}
 C_{x,x}=C_{y,y}=C_{z,z}=-1, & {\text { if $S=0$}}.
\end{eqnarray}
 
 In order to find spin-transfer coefficients, one should consider
 the following cross section
 \begin{equation}
 \label{s0kij}
  d\sigma({\bf p}_1,{\bf p}_3)= \Phi\sum_{\mu_2,\mu_4}
|T_{\mu_1\,\mu_2}^{\mu_3\,\mu_4}|^2.
  \end{equation}
 The polarization vector ${\bf p}_1$ of the  {\it 1}-st particle in
 the right side of Eq.(\ref{s0kij}) can be found only
 in the following sum
  \begin{eqnarray}
 \label{s0kij2}
 \sum_{\mu_2}(\frac{1}{2}\mu_1\frac{1}{2}\mu_2|00)^2=
\frac{1}{2}\sum_{\mu_2} (\chi_{\mu_1}^{+} {i\sigma_y}
 \chi_{\mu_2}^{(T)+})(\chi_{\mu_2}^{T}{(-i\sigma_y)}
\chi_{\mu_1})= \ \ \ \ \nonumber \\ 
=\frac{1}{4} Sp(1+{\bfg \sigma} \cdot
 {\bf p}_1)=\frac{1}{2}.\ \ \ \ \
 \end{eqnarray}
 Since the vector ${\bf p}_1$ is absent actually in the right hand
 side of Eq.
 (\ref{s0kij2}), one should conclude that the all polarization 
 transfer coefficients are zero for the spin-singlet initial state:
 $K_i^j=0\, (i,j=x,y,z)$.  The obtained results for $C_{i,j}$ and $K_i^j$
 are valid for any values of the spins  $j_3$ and $j_4$,
 both of them being integer or half-integer.
 
 \subsection  {The spin-triplet initial state.}

  For $S=1$ and $ M=0$,  Eq. (\ref{polarsec})   can be written as
 \begin{eqnarray}
 \label{m0}
 d\sigma_{M=0}({\bf p}_1,{\bf p}_2)=
\frac{\Phi}{16\,\pi}(1+{\bf
    p}_1\cdot {\bf p}_2-2p_{1z}p_{2z})\,\times \ \ \ \ \ \nonumber \\
 \sum_J|\sqrt{J}\,
  a_J^{J-1}-\sqrt{J+1}\,a_J^{J+1}|^2. \ \ \ \ \ \ 
\end{eqnarray}
 We obtain  this formula from Eq. (\ref{polarsec})   using  
Eq. (\ref{s1}) and 
the  following formulae for the Clebsch-Gordan coefficients:
$(1\,0\, J\,0|J\,0)=0,(1\,0\, J-1\,0|J\,0)=\sqrt{J/(2J-1)}$, 
$(1\,0\, J+1\,0|J\,0)=-\sqrt{(J+1)/(2J+3)}$.
 In order to  simplify the notations,  we omit in Eq. (\ref{m0})
 and below
  the superscript $S=1$ in $a_J^{L\,S}$. The sum over the projections
  $M= +1$ and $M=-1$ into right-hand side of  Eq. (\ref{polarsec})
gives 
\begin{eqnarray}
\label{mpm1}
 d\sigma_{M=\pm1}({\bf p}_1,{\bf p}_2)=
\frac{\Phi}{16\,\pi}(1+p_{1z}p_{2z})\times  \ \ \ \ \ 
\nonumber \\
\times \begin{cases}
\sum_J\,
|\sqrt{J}\, a_J^{J+1}+\sqrt{J+1}\,a_J^{J-1}|^2, &
 {\text {if \, $(-1)^{J+1}=\pi$,}}\\
\sum_{J}\,(2J+1)\,|a_J^J|^2, & {\text {if\, $(-1)^{J}=\pi$}}.
\end{cases}
\end{eqnarray}
 Here we used the following relations: $ (11\,J\,-1|J\,0)=\frac{1}{\sqrt{2}}$,
$(11\,J\,-1|J-1\,0)=\sqrt{J+1}/\sqrt{2(2J+1)}$,
$(11\,J\,-1|J+1\,0)=\sqrt{J}/\sqrt{2(2J+1)}$.
Using Eqs.(\ref{m0}), (\ref{mpm1}),
 one can make summation
 over $M$ in Eq. (\ref{polarsec}) and then present
 the spin-triplet polarized cross section 
 in the  following standard form \cite{ohlsen} 
\begin{equation}
\label{standard}
 d\sigma({\bf p}_1,{\bf p}_2)= d\sigma_0\,
(1 + C_{x,x}\,p_{1x}\,p_{2x}+ C_{y,y}\,p_{1y}p_{2y}+
C_{z,z}\,p_{1z}\,p_{2z}),
\end{equation}
where the spin-spin correlation parameters are given as
\begin{eqnarray}
\label{cyys1}
  C_{x,x}=C_{y,y}= \frac{\sum_J|\sqrt{J}\,
  a_J^{J-1}-\sqrt{J+1}\,a_J^{J+1}|^2} {\sum_{J\,L}(2J+1)\,|a_J^L|^2},
  & {\text {\,\, \,\, if $S=1$},} \\
\label{czzs1}
 C_{z,z}=1-2\,C_{y,y}.
\end{eqnarray}
 As seen from Eq. (\ref{cyys1}), the spin-spin correlation parameters
 are non-negative for transversal polarization. 
 One can see from Eq. (\ref{cyys1}) that the diagonal term $a_J^J$
 does not contribute into the numerator of $C_{x,x}=C_{y,y}$. 
 The obtained results for $C_{i,j}$ 
 are valid for any values of the spins  $j_3$ and $j_4$,
 both of them being integer or half-integer.
 Eqs. (\ref{cyys1}) and (\ref{czzs1}) are very simple for applications.
 Explicite formulae for different spins of  the $\Theta^+$ will be given
 in the next section.

 Considering the sum
 $\sum_{\mu_2}(\frac{1}{2}\mu_1\frac{1}{2}\mu_2|1M)
(\frac{1}{2}\mu_1\frac{1}{2}\mu_2|1M')$, one can find that
this sum  explicitly contains the polarization
 vector ${\bf p}_1$. Therefore, in contrast to the case
 of $S=0$, the spin-triplet initial state $S=1$ allows 
 a  non-zero polarization transfer in this reaction. 
 In order to get  the spin-transfer coefficients we use below
 a general method developed in Ref. \cite{bilenky}.

\section { General method} 

 According to Ref. \cite{bilenky},
  the amplitude in Eq. (\ref{tfi}) can be written as 
\begin{equation}
\label{operatortfi}
 T_{\mu_1\,\mu_2}^{\mu_3\,\mu_4}=
\chi_{j_3\,\mu_3}^+\chi_{j_4\,\mu_4}^+ {\hat F}\chi_{j_1\,\mu_1}\,
\chi_{j_2\,\mu_2},
\end{equation}
where $ {\hat F}$ is an operator acting on the spin states  of the
initial and final particles.
 This operator can be written as
\begin{equation}
\label{f}
{\hat F}= \sum_{m_1\,m_2\,m_3\,m_4}
T_{m_1\,m_2}^{m_3\,m_4}\, \chi_{j_1\,m_1}^+(1) \chi_{j_2\,m_2}^+(2)
\chi_{j_3\,m_3}(3)\,\chi_{j_4\,m_4}(4),
\end{equation} 
where $\chi_{j_k\, m_k} (k)$ is the spin function of the $k$th
particle with the spin $j_k$ and z-projection $m_k$ and 
$ T_{m_1\,m_2}^{m_3\,m_4}$  is defined by Eq. (\ref{tfi}).
The operator ${\hat F}$ is
 normalized to the unpolarized cross section as
\begin{equation}
\label{trff}
d\sigma_0 =\frac{\Phi}{(2j_1+1)\,(2j_2+1)} Sp FF^+.
\end{equation}
For the spin-triplet case $(S=1)$ with $j_1=j_2=j_3=\frac{1}{2}$ 
 one can write
\begin{eqnarray}
\label{spurff}
4\pi SpFF^+ = \,\,\,\,\,\,\,\,\,\,\,\,\,  \nonumber \\ 
= \begin{cases}
(2J_p+1) |a_{J_p}^{J_p}|^2+(2J_m+1)\bigl (|a_{J_m}^{J_m+1}|^2+ 
|a_{J_m}^{J_m-1}|^2\bigr )
 &{\text {if  $(-1)^{J_p}=\pi$,}}\\ 
(2J_m+1) |a_{J_m}^{J_m}|^2+(2J_p+1)\bigl (|a_{J_p}^{J_p+1}|^2+ 
|a_{J_p}^{J_p-1}|^2 \bigr )
 &{\text {if  $(-1)^{J_m}=\pi$}}
\end{cases}
\end{eqnarray}
 for $j_4>\frac{1}{2}$ and
\begin{eqnarray}
\label{spurffmin}
4\pi SpFF^+ =  
\begin{cases}
3 |a_{1}^{1}|^2+|a_{0}^{1}|^2 
 &{\text {if  $\pi=-1$,}}\\ 
3 |a_{1}^{2}|^2+3|a_{1}^{0}|^2 
 &{\text {if  $\pi= +1$}}
\end{cases}
\end{eqnarray}
 for $j_4=\frac{1}{2}$. 

\subsection {Spin- transfer coefficients.}

  The spin-transfer coefficient, describing the polarization transfer
 from the
{\it 1}st  particle to the  {\it 3}-rd one in the reaction
${\vec j_1}+j_2\to{\vec j_3}+j_4$, is given by the following formula
\cite{bilenky}
\begin{equation}
\label{kijff}
 K_\lambda^\kappa= \frac{Sp F\,{\hat j}_\lambda(1)\,F^+\,{\hat j}_\kappa(3)}
{Sp F\,F^+\,j_1j_3},
 \end{equation}
 where ${\hat j}_\lambda(1)$ and ${\hat j}_\kappa(3)$ are
 the spherical components of the spin operators of the particles
 1 and 3 ($\lambda,\kappa=0,\,\pm 1$), respectively.
 For  arbitrary spins of all participating particles $j_i$ ($i=1,\dots,4$)
  we found from
 Eqs. (\ref{tfi}), (\ref{operatortfi}) and (\ref{kijff})
 the spin transfer
 coefficient in the following general form (for details, see Appendix):
\begin{eqnarray}
\label{kijgen}
 K_\lambda^{\kappa}(1\to 3)\,Sp FF^+=
\delta_{\lambda,-\kappa}\,\frac{1}{4\pi}\, \sqrt{j_1^{-1}j_3^{-1}
(j_1+1)(2j_1+1)(j_3+1)(2j_3+1)}\times \nonumber \\
\times \sum_{S\,S'\,J\,J'\,L\,L'\, J_0}
\sqrt{(2L+1)\,(2L'+1)\,}\times\nonumber \\  
\times \sqrt{(2S+1)\,(2S'+1)}
(2J+1)\,(2J'+1)\,(-1)^{j_1+j_2+j_3+j_4 + S'+J'+L+1}\times  \nonumber \\
\times (1\, -\lambda \, 1\, \lambda |J_0\, 0)
(L'\, 0 \, L\, 0\,|J_0 0)\times \ \ \  \nonumber \\
 \left \{\begin{array}{ccc}
j_1 & j_2 & S \\
S' & 1 & j_1  
\end{array} \right \}
\left \{\begin{array}{ccc}
j_3 & j_4 & J' \\
J & 1 & j_3  
\end{array} \right \}
\left \{\begin{array}{ccc}
J & S & L \\
J' & S' & L' \\
1 & 1 &  J_0 
\end{array} \right \} a_J^{L\,S} (a_{J'}^{L'\,S'})^* .\ \ \ \ \ \  
\end{eqnarray} 
 Here we used the standard notations for the 6j-and 9j-symbols
 \cite{Varschalovich}. In Eq. (\ref{kijgen}) the
 intermediate angular momentum $J_0$ is even: 
$J_0=0$ and 2. The value $J_0=1$ is excluded
because the Clebsch-Gordan coefficient $(L'\, 0\, L\, 0\,| J_0 \,0)$ equals
 zero for $L'+L+J_0=odd$ and, furthermore, $L'+L$ is even due to P-parity
 conservation. Taking into account this fact, 
   one can find from Eq.(\ref{kijgen}) the
 following relations (for details, see Appendix):
\begin{eqnarray}
\label{k}
K_{+1}^{-1}=K_{-1}^{+1}= -K_x^x=-K_y^y, \, \,  K_0^0=K_z^z,\,\,
K_z^x=K_z^y=K_x^z=K_y^z=0,
\end{eqnarray}
  and $K_i^j=0$ at $i\not =j$, where $i,j=x,y,z$.
  From Eq. (\ref{kijgen}) we also find   that
 there is no polarization transfer 
 ($K_i^j=0$, $i,j=x,y,z)$ 
 for $S=S'=0$ in accordance with the 
 discussion given in the section \ref{sect31}.
 These coefficients are also equal to  zero   for $J=J'=0$:
\begin{eqnarray}
\label{kijsinglet}
 K_i^j=0 \,\,\, (i,j=x,y,z),  & {\text {\, \, \, \, if $S=S'=0$ or $J=J'=0$}.}
 \end{eqnarray}
 For the spin-triplet transitions $S=S'=1$, we find from
 Eq. (\ref{kijgen}) that $K_x^x=K_y^y\not=0$ and  $K_0^0\equiv K_z^z\not =0$.
  Eq. (\ref{kijgen}) is a generalization of the formula given by Eq.(5) 
 in Ref. \cite{uzikov2} for the particular case of $j_1=j_3=\frac{1}{2}$.

 Eq. (\ref{kijgen}) is rather
 complicated due to its general applicability. For particular case of
 $j_1=j_2=j_3=\frac{1}{2}$,   we present below explicit formulae
 for $K_i^j$
 for $j_4=\frac{1}{2},\, \frac{3}{2}$ and $\frac{5}{2}$.

\subsubsection{ $j_4= \frac{1}{2}$}

 For the total isospin $T=0$ and
 the P-parity  $\pi=+1$
 one has  $S=1$. For this case Eq. (\ref{kijgen}) gives
 (using the notation $a_J^{L\,1}=a_J^L$)
\begin{eqnarray}
\label{onehalfplusx}
K_y^y=
\frac{|\sqrt{2}\,a_1^0+a_1^2|^2-3\,Re\,(\sqrt{2}\,a_1^0+a_1^2){a_1^2}^*}
{3\,(|a_1^0|^2+|a_1^2|^2)}, \\
\label{onehalfplusz}
K_z^z=
\frac{|\sqrt{2}\,a_1^0+a_1^2|^2}
{3\,(|a_1^0|^2+|a_1^2|^2)}. 
\end{eqnarray}
 For   $T=1$ and
 $\pi=-1$ one has   $S=1$. In this case  Eq. (\ref{kijgen})
 gives 
\begin{eqnarray}
\label{onehalfminusx}
 K_y^y=\frac{\sqrt{6}\, Re\,a_0^1\,{a_1^1}^*}
{|a_0^1|^2+3|a_1^1|^2},\\
 \label{onehalfminusz}
 K_z^z=\frac{3|a_1^1|^2}{|a_0^1|^2+3|a_1^1|^2}, 
\end{eqnarray}
 which  coincide (except for notations)  with those
 obtained recently in Ref. \cite{rekalo2} in the $\bfg
 \sigma$-representation for the amplitude.

\subsubsection {$j_4=\frac{3}{2}$}

 For higher  spins of the
 {\it 4}th particle $j_4\geq \frac{3}{2}$, Eq. (\ref{kijgen})
 also gives non-zero coefficients 
 $K_x^x$ and  $K_z^z$ for $S=1$.  So, for  $j_4=\frac{3}{2}$ 
 we find
\begin{eqnarray}
\label{threehalfminus}
4\pi\, SpFF^+\,K_y^y={ 3|a_2^1|^2 - 3|a_2^3|^2+
Re\left (\sqrt{3}a_1^1\,{a_2^{1}}^* -\sqrt{\frac{9}{2}} a_1^1{a_2^3}^*
 -\sqrt{\frac{3}{2}} a_2^3\,{a_2^1}^*\right )},\nonumber \\
4\pi\, SpFF^+\, K_z^z= {\frac{3}{2}(|a_2^1|^2-|a_1^1|^2) +|a_2^3|^2
 +Re \left (\sqrt{18}
  a_1^1{a_2^3}^*+\sqrt{27}a_1^1\,{a_2^1}^*+
\sqrt{6}
  a_2^3\,{a_2^1}^* \right )},\nonumber \\
 \end{eqnarray}
if $T=1$   and $\pi=-1$,  and
\begin{eqnarray}
\label{threehalfplus}
4\pi\, SpFF^+\,  K_y^y={ |a_1^2|^2 - |a_1^0|^2 +
Re\left (\frac{\sqrt{2}}{2} a_1^2{a_1^0}^*
-\sqrt{\frac{15}{2}}a_2^2\,{a_1^{0}}^* +
 \sqrt{15}a_2^2\,{a_1^2}^* \right )},\nonumber \\
4\pi\, SpFF^+\, K_z^z= 
{\frac{5}{2}|a_2^2|^2-|a_1^0|^2-\frac{1}{2}|a_1^2|^2 
+Re\left (\sqrt{30}a_2^2\,{a_1^{0}}^* -\sqrt{2} a_1^2{a_1^0}^*+
 \sqrt{15}  a_2^2\,{a_1^2}^*\right )},\nonumber \\
 \end{eqnarray}
if $T=0$  $\pi=+1$. The factors $ SpFF^+$ in
 Eqs. (\ref{threehalfminus}) and (\ref{threehalfplus})
  are given by Eq. (\ref{spurff}).
\subsubsection{$j_4=\frac{5}{2}$}

 We consider here   the case of $j_4=\frac{5}{2}$, because there is 
 an experimental indication for this value of the spin of the $\Theta^+(1540)$
 \cite{troyan}. 
 For $j_4=\frac{5}{2}$  we find
\begin{eqnarray}
\label{fivehalfminus}
4\pi\, SpFF^+\,K_y^y={2(|a_2^3|^2 - |a_2^1|^2 )+
Re \left (2\sqrt{7} a_3^3{a_2^3}^* -\frac{2\sqrt{42}}{3}a_3^3{a_2^1}^*+
\frac{\sqrt{6}}{3}a_2^3\,{a_2^1}^*\right )},\nonumber \\
4\pi\, SpFF^+\,K_z^z=
{\frac{7}{3}|a_3^3|^2 -|a_2^1|^2-\frac{2}{3}|a_2^3|^2 +
Re \left (\frac{8\sqrt{7}}{3} a_3^3{a_2^3}^*+
\frac{4\sqrt{42}}{3}a_3^3{a_2^1}^*-
\frac{2\sqrt{6}}{3}a_2^3\,{a_2^1}^*\right )},\nonumber \\
 \end{eqnarray}
if $T=1$ and $\pi=-1$, and 
\begin{eqnarray}
\label{fivehalfplus}
4\pi\, SpFF^+\,  K_y^y={4(|a_3^2|^2 - |a_3^4|^2) -
Re \left (\frac{2\sqrt{30}}{3}a_2^2{a_3^4}^*-\sqrt{10}a_2^2\,{a_3^2}^*+
\frac{2\sqrt{3}}{3}a_3^4\,{a_3^2}^*\right )},\nonumber \\
4\pi\, SpFF^+\,K_z^z={-\frac{5}{3}|a_2^2|^2+\frac{4}{3}|a_3^2|^2 +|a_3^4|^2+
Re \left (\frac{4\sqrt{30}}{3} a_2^2{a_3^4}^*+\frac{8\sqrt{10}}{3}
a_2^2\,{a_3^2}^* +
\frac{4\sqrt{3}}{3}a_3^4\,{a_3^2}^* \right )
}\nonumber \\
 \end{eqnarray}
  if $T=0$ and $\pi=+1$. 

\subsection {Spin-spin correlation coefficients.}
In this section we consider spin-spin correlations in the initial and 
final states of the reaction $1+2\to 3+4$.
 \subsection{Spin-spin correlation in the initial state
 ${\vec j_1}+{\vec j_2}\to j_3+j_4$}

  The initial spin-spin correlation coefficient is defined as \cite{ohlsen}
\begin{equation}
\label{clk}
 C_{\lambda\,,\kappa}=\frac{Sp F\, {\hat j}_\lambda(1)\,
{\hat j}_\kappa(2)\,F^+}
 {Sp F\,F^+\, j_1 j_2},
 \end{equation}
where ${\hat j}_\alpha(i)\, (\alpha=0,\pm 1)$ is the spin operator
 of the {\it i}-th particle. Using Eqs.(\ref{tfi}) and (\ref{clk}), 
we find for arbitrary spins $j_i$ ($i=1,\dots, 4)$
\begin{eqnarray}
\label{cijgen}
C_{\lambda,\kappa}\, Sp\, FF^+=
\delta_{\lambda,-\kappa}\,
\frac{ 1}{4\pi}\sqrt{j_1^{-1}\,
j_2^{-1}\,(j_1+1)(2j_1+1)(j_2+1)(2j_2+1)}\times
 \nonumber \\
\times \sum_{S\,S'\,J}(-1)^{S+J} (2J+1)
 \sqrt{(2S+1)(2S'+1)}\times \nonumber \\
\times \sum_{L\,L'\,J_0}(-1)^{L}\sqrt{(2J_0+1)(2L'+1)}\times \nonumber \\
\times (1\lambda \,1-\lambda\,|J_0\,0)\,(J_0\,0 \,L'\, 0\,|L\,0)\times 
\nonumber \\
\hspace{-2cm} \times  \left \{\begin{array}{ccc}
S' & S & J_0 \\
L & L' & J  
\end{array} \right \}
\left \{\begin{array}{ccc}
S^\prime & j_1 & j_2 \\
S & j_1 &  j_2 \\
J_0 & 1 &  1 
\end{array} \right \}
 a_J^{L\,S} (a_{J'}^{L'\,S'})^*.\, \,
\end{eqnarray}
Using relations similar to those, which were employed for deriving of
 Eq.(\ref{k})(see Appendix),
 we find from Eq. (\ref{cijgen}) the following relations:
\begin{equation}
\label{c}
C_{+1,-1}=C_{-1,+1}=-C_{x,x}=-C_{y,y}\not =0,
\end{equation}
 and  $C_{0,0}=C_{z,z}\not=0$,
whereas 
\begin{eqnarray}
C_{i,j}=0 & {\text { \,\,\,\,    if $i\not =j,\,  (i,j=x,y,z)$}}.
\end{eqnarray} 

 One can see  from
 Eq. (\ref{clk}) that the  following relation holds independently on the
 production mechanisms:
\begin{eqnarray}
\label{summacii}
 \Sigma=C_{x,x}+C_{y,y}+C_{z,z}=
{\hat {\bf j }}(1)\cdot {\hat {\bf j }}(2)/(j_1\,j_2).
\end{eqnarray}
 Therefore, $\Sigma$ is
 fixed by the spin $S$. For $j_1=j_2=\frac{1}{2}$, one has
 $\Sigma=-{3}$
 for  $S=0$  and $\Sigma=+{1}$ for $S=1$
 in accordance with the above results given in 
 Eqs. (\ref{cyys1}), (\ref{czzs1}) and (\ref{s0sec}).
 From  Eq. (\ref{cijgen}) one can find that
  $C_{x,x}=C_{y,y}=C_{z,z}=-1$ for $S=S'=0$.
 For  $S=S^\prime=1$, Eq. (\ref{cijgen}) is a generalization of
  Eqs.({\ref{cyys1}) and  (\ref{czzs1}), derived for the
 particular case $j_1=j_2=\frac{1}{2}$ 
 
%

 Using Eq.(\ref{cijgen}) one can write explicite formulae for $C_{i,j}$,
 which  coincide with those obtained from the more simple formula in
 Eq.(\ref{cyys1}). We present below the formulae for $C_{y,y}$ for
 different $j_4$ at  
 $j_1=j_2=j_3=\frac{1}{2}$ and $S=1$. The coefficient $C_{z,z}$ can be
  found as $C_{z,z}=1-2C_{y,y}$.
 
\subsubsection{ $j_4=\frac{1}{2}$.}
\begin{eqnarray}
\label{cij12}
 C_{y,y}=\frac{|a_0^1|^2}{|a_0^1|^2+3|a_1^1|^2 },
 & {\text {\, \, \, \, if $\pi=-1$ and  $T=1$};} \\
\label{cij12pl}
 C_{y,y}=\frac{|a_1^0-\sqrt{2}a_1^2|^2}{3|a_1^0|^2+3|a_1^2|^2},
 & {\text {\, \, \, \, if $\pi=+1$ and  $T=0$}.}
\end{eqnarray}
\subsubsection{ $j_4=\frac{3}{2}$.}
\begin{eqnarray}
\label{cij32}
C_{y,y}=\frac {|\sqrt{2}\, a_2^1-\sqrt{3}\, a_2^3|^2}
{3|a_1^1|^2+5|a_2^1|^2+5|a_2^3|^2},
 & {\text {\, \, \, \, if $\pi=-1$ and  $T=1$};}\nonumber \\
C_{y,y}=\frac {|a_1^0-\sqrt{2}\, a_1^2|^2}
{3|a_1^0|^2+3|a_1^2|^2+5|a_2^2|^2},
 & {\text {\, \, \, \, if $\pi=+1$ and  $T=0$}.}
\end{eqnarray}
\subsubsection{ $j_4=\frac{5}{2}$.}
\begin{eqnarray}
\label{cij52}
C_{y,y}=\frac {|\sqrt{2}\, a_2^1-\sqrt{3}\, a_2^3|^2}
{5|a_2^1|^2+5|a_2^3|^2+7|a_3^3|^2},
 & {\text {\, \, \, \, if $\pi=-1$ and  $T=1$};}\nonumber \\
C_{y,y}=\frac {|\sqrt{3}a_3^2-{2}\, a_3^4|^2}
{5|a_2^2|^2+7|a_3^2|^2+7|a_3^4|^2},
 & {\text {\, \, \, \, if $\pi=+1$ and  $T=0$}.}
\end{eqnarray}

\subsection{Spin-spin correlation in the final state
 $j_1+j_2\to {\vec j_3}+{\vec j_4}$ }

 By analogy with Eq. (\ref{clk}),
 for the spin-spin correlation in the final state, we can use the following
 definition
\begin{equation}
\label{clkf}
 C_{\lambda\,,\kappa}^f=\frac{Sp F^+\, {\hat j}_\lambda(3)\,
{\hat j}_\kappa(4)\,F}
 {Sp F\,F^+\,j_3\,j_4}.
 \end{equation}
 We should note that in comparison with Eq. (\ref{clk}) the right
 hand side of Eq. (\ref{clkf})  looks as the spin-spin correlation
 in the initial state of the inverse  reaction $3+4\to 1+2$.
 For  arbitrary spins $j_3$ and $j_4$ we find from Eq. (\ref{clkf})

\begin{eqnarray}
\label{clkfinal}
C_{\lambda,\kappa}^f\, Sp\, FF^+=
\delta_{\lambda,-\kappa}\,
\frac{ 1}{4\pi}\sqrt{j_3^{-1}\,j_4^{-1}\,(j_3+1)(2j_3+1)\,(j_4+1)
(2j_4+1)}\times
 \nonumber \\
\times \sum_{J\,J'\,S}(-1)^{S+J} (2J+1)(2J'+1)
\times \nonumber \\
\times \sum_{L\,L'\,J_0}(-1)^{L}\sqrt{(2J_0+1)(2L'+1)}\times \nonumber \\
\times (1\lambda \,1-\lambda\,|J_0\,0)\,(J_0\,0 \,L'\, 0\,|L\,0)\times 
\nonumber \\
\hspace{-2cm} \times  \left \{\begin{array}{ccc}
J' & S & L' \\
L & J_0 & J  
\end{array} \right \}
\left \{\begin{array}{ccc}
J & j_3 & j_4 \\
J' & j_3 &  j_4 \\
J_0 & 1 &  1 
\end{array} \right \}
 a_J^{L\,S} (a_{J'}^{L'\,S'})^*.\, \,
\end{eqnarray}
 Here $J_0=0$ and $2$. Others angular momenta $J,J',S,L,L'$ are defined
 in the section \ref{sect1}.This formula is valid for arbitrary values
 of the spins $j_i$ 
 being integer or half-integer.
 We found from Eq. (\ref{clkfinal}) the following relations:
 $C_{+1,-1}^f=C_{-1,+1}^f=-C_{x,x}^f=-C_{y,y}^f\not =0$,  $C_{0,0}^f
 =C_{z,z}^f\not=0$,
 whereas $C_{i,j}^f=0$ at $i\not =j$ ($i,j=x,y,z$). 
  Below we derive explicite formulae 
 for $C_{\lambda,\kappa}^f$ for $j_4=\frac{1}{2}$ and $ \frac{3}{2}$, 
 assuming $j_1=j_2=j_3=\frac{1}{2}$.

 One can see from Eq. (\ref{clkf}) that the sum of the diagonal terms
 $C_{i,j}$ is determined by the total spin 
${\hat {\bf J}}={\hat {\bf j}}(3)+{\hat {\bf j}}(4)$
 as
 \begin{eqnarray}
 \label{sumcijf}
 \Sigma^f=C_{x,x}^f+C_{y,y}^f+C_{z,z}^f=\nonumber \\
 ={\hat {\bf j}}(3)\cdot {\hat {\bf j}}(4)/(j_3j_4)=\frac{1}{2j_3j_4}\left 
[J(J+1)-j_3(j_3+1)-j_4(j_4+1)
 \right].
 \end{eqnarray}
 We will use below Eq.(\ref{sumcijf})  as a  check for the derived
 formulae for $C_{i,j}^f$.

\subsubsection{ The case of $j_4=\frac{1}{2}$}

 One can find from Eq.({\ref{clkfinal})
\begin{eqnarray}
\label{cft1pmin}
C_{x,x}^f=C_{y,y}^f=C_{z,z}^f=-1, & {\,\,\, \text {if $\pi=+1$ and $T=1$}}
\end{eqnarray}
and 
\begin{eqnarray}
\label{cft1pplus}
C_{y,y}^f=-\frac{|a_0^1|^2}{|a_0^1|^2+3|a_1^1|^2}, 
C_{z,z}^f=\frac{3|a_1^1|^2-|a_0^1|^2}{|a_0^1|^2+3|a_1^1|^2},
& {\,\,\, \text {if $\pi=-1$ and $T=1$}} \nonumber \\
\Sigma^f=\frac{3|a_1^1|^2-3|a_0^1|^2}{|a_0^1|^2+3|a_1^1|^2}=
\begin{cases}
+1,  {\text{  $J=1$,}} \\ 
-3,  {\text {$J=0$}}.
\end{cases} 
\end{eqnarray}

 The final
 spin-spin correlation coefficients for $\pi=+1$  given by Eq. (\ref{cft1pmin})
 differ from those for $\pi=-1$ given by Eq. (\ref{cft1pplus}).
 The only exception is the point where $a_1^1=0$ (but $a_0^1\not= 0$).

For the case of $T=0$ one has  
\begin{eqnarray}
\label{cft0pmin}
C_{x,x}^f= C_{y,y}^f=+1,\,\, C_{z,z}^f=-1, & {\text { \,\,\, if $\pi=-1$ and
 $T=0$.}}
\end{eqnarray}
and 
\begin{eqnarray}
\label{cft0pplus}
C_{y,y}^f=\frac{|a_1^0-\sqrt{2}a_1^2|^2}
{3(|a_1^0|^2+|a_1^2|^2)},\,\, C_{z,z}^f=1-2C_{y,y}, &
 {\text { \,\,\,\,    if $\pi=+1$ and $T=0$}}
\end{eqnarray}
Eq. (\ref{cft0pplus}) coincides with Eq. (\ref{cij12}).
 One can see from Eqs. (\ref{cft0pmin}) and (\ref{cft0pplus}),
 the final spin-spin correlation coefficients are different in absolute value
 for positive  and negative parity $\pi$. Only exception is the point
 $a_1^2=-\sqrt{2}a_1^0$ at which the coefficients $C_{j,j}^f$  for $\pi=+1$
 are identical to those for $\pi=-1$ and have  a maximal absolute value.
One can see, Eq. (\ref{cft0pplus}) coincides with Eq. (\ref{cij12pl}):
$ C_{x,x}^f=C_{y,y}^f=C_{x,x}^i=C_{y,y}^i$.
 
\subsubsection{ The case of $j_4=\frac{3}{2}$}

For $T=1$ and $\pi=-1$ we find
\begin{eqnarray}
\label{ckf32t1minuss}
C_{y,y}^f=\frac{1}{6}\frac{20(|a_2^1|^2+|a_2^3|^2)
- |3a_1^1+\sqrt{3}a_2^1+\sqrt{2}a_2^3|^2}
{3|a_1^1|^2+5|a_2^1|^2+5|a_2^3|^2},\nonumber \\
C_{z,z}^f=\frac{1}{3}\frac{|3a_1^1+\sqrt{3}a_2^1+\sqrt{2}a_2^3|^2-
5(3|a_1^1+|a_2^1|^2+|a_2^3|^2)}
{3|a_1^1|^2+5|a_2^1|^2+5|a_2^3|^2}.
\end{eqnarray}
\begin{eqnarray}
\label{ckf32t1minuszig}
\Sigma^f=\frac{1}{3}\frac{-15|a_1^1|^2+15|a_2^1|^2+|a_2^3|^2}
{3|a_1^1|^2+5|a_2^1|^2+5|a_2^3|^2}=
\begin{cases}
-\frac{5}{3},  {\text{  $J=1$,}} \\ 
+1,  {\text {  $J=2$}}.
\end{cases} 
\end{eqnarray}
For $T=1$ and $\pi=+1$ one has 
\begin{eqnarray}
\label{cf32t1plus}
C_{y,y}^f=+\frac{2}{3}, C_{z,z}^f=-\frac{1}{3}.
\end{eqnarray}
 Let us consider the case of $T=0$. For $\pi=+1$ we find
\begin{eqnarray}
\label{ckf32t0plus}
C_{y,y}^f=\frac{\frac{15}{6}|a_2^2|^2-\frac{11}{6}|a_1^2|^2
-\frac{10}{6}|a_1^0|^2-Re\left (\frac{\sqrt{15}}{3}\,a_2^2\,{a_1^2}^*+
\frac{\sqrt{30}}{3}\,a_2^2{a_1^0}^*-
\frac{\sqrt{2}}{3}\, a_1^2{a_1^0}^*\right )}
{3|a_1^0|^2+3|a_1^2|^2+5|a_2^2|^2},\nonumber \\
C_{z,z}^f=\frac{-\frac{4}{3}|a_1^2|^2-\frac{5}{3}|a_1^0|^2
+2Re \left (\frac{\sqrt{15}}{3}\,a_2^2\,{a_1^2}^*+
\frac{\sqrt{30}}{3}\,a_2^2{a_1^0}^*-
\frac{\sqrt{2}}{3}\, a_1^2{a_1^0}^*\right )}
{3|a_1^0|^2+3|a_1^2|^2+5|a_2^2|^2},\nonumber \\
\Sigma^f=\frac{5[|a_2^2|^2-|a_1^2|^2-|a_1^0|^2]}
{3|a_1^0|^2+3|a_1^2|^2+5|a_2^2|^2}=
\begin{cases}
-\frac{5}{3},  {\text{  $J=1$,}} \\ 
+1,  {\text {  $J=2$}}. \ \ \ \ \
\end{cases} 
\end{eqnarray}

For $T=0$ and $\pi=-1$ one can find
\begin{eqnarray}
\label{cf32t0minus}
C_{x,x}^f=C_{y,y}^f=-\frac{2}{3}, C_{z,z}^f=-\frac{1}{3}
\end{eqnarray}

\subsubsection{ The case of $j_4=\frac{5}{2}$}

 For $T=1$ and $\pi=-1$
\begin{eqnarray}
\label{ckf52t1minusx}
C_{y,y}^f=\frac{1}{15}\frac{56|a_3^3|^2-43|a_2^3|^2-42|a_2^1|^2-
 2Re\left(4\sqrt{7}a_3^3{a_2^3}^*+2\sqrt{42}a_3^3{a_2^1}^*-
\sqrt{6}a_2^1{a_2^3}^*\right)}
{5|a_2^1|^2+5|a_2^3|^2+7|a_3^3|^2},\nonumber \\
C_{z,z}^f=\frac{1}{15}\frac{-7|a_3^3|^2-{19}|a_2^3|^2-21|a_2^1|^2+
4Re\left[4\sqrt{7}a_3^3{a_2^3}^*+2\sqrt{42}a_3^3{a_2^1}^*-
\sqrt{6}a_2^1{a_2^3}^*\right ]}
{5|a_2^1|^2+5|a_2^3|^2+7|a_3^3|^2}, \ \ \ \nonumber \\
\Sigma^f=7
\frac{\left (|a_3^3|^2-|a_2^1|^2-|a_2^3|^2\right )}
{5|a_2^1|^2+5|a_2^3|^2+7|a_3^3|^2}=
\begin{cases}
-\frac{7}{5},  {\text{  $J=2$,}} \\ 
+1,  {\text {  $J=3$}}. \ \ \ \ \ \ 
\end{cases} 
\end{eqnarray}
For the singlet initial  state at $T=1$ and $\pi=+1$ we find
\begin{eqnarray}
\label{cf352t1pl}
C_{x,x}^f=C_{y,y}^f=-\frac{3}{5},\, C_{z,z}^f=-\frac{1}{5}.
\end{eqnarray}

For $T=0$ and $\pi=+1$ we find
\begin{eqnarray}
\label{ckf52t0plus}
C_{y,y}^f=\frac{1}{15}\frac{-40|a_2^2|^2+60|a_3^4|^2+59|a_3^2|^2-
4 Re\left[\sqrt{30}a_2^2{a_3^4}^*+2\sqrt{10}a_2^2{a_3^2}^*+
\sqrt{3}a_3^4{a_3^2}^*\right ]}
{5|a_2^2|^2+7|a_3^4|^2+7|a_3^2|^2},\nonumber \\
C_{z,z}^f=\frac{1}{15}\frac{-25|a_2^2|^2-13|a_3^2|^2-15|a_3^4|^2+
8 Re\left[\sqrt{30}a_2^2{a_3^4}^*+2\sqrt{10}a_2^2{a_3^2}^*+
\sqrt{3}a_3^4{a_3^2}^*\right ]}
{5|a_2^2|^2+7|a_3^4|^2+7|a_3^2|^2}, \nonumber \\
\Sigma^f=7
\frac{\left (-|a_2^2|^2+|a_3^2|^2+|a_3^4|^2\right )}
{5|a_2^2|^2+7|a_3^4|^2+7|a_3^2|^2}   =
\begin{cases}
-\frac{7}{5},  {\text{  $J=2$,}} \\ 
+1,  {\text {  $J=3$}}. \ \ \ \ \ \ 
\end{cases} 
\end{eqnarray}
For the singlet initial  state at $T=0$ and $\pi=-1$ one can find
\begin{eqnarray}
\label{cf352t0min}
C_{x,x}^f=C_{y,y}^f=\frac{3}{5},\, C_{z,z}^f=-\frac{1}{5}.
\end{eqnarray}

\subsection{Spin-tensor correlation in the final state
            induced by polarized beam}

 The spin-density matrix of the particle with the spin $j$ is determined
 by $(2j+1)(2j+1)$ tensor momenta, $t_{JM}$. Here $J$ is   the rank
($J=0,1,\dots, 2J$)  and $M$ is the magnetic quantum number 
   ($M=-J, -J+1, \dots, +J$). Let us define the
 spin-transfer coefficient  for induced tensor polarization as
\begin{equation}
\label{ttt}
K_{J_1\,M_1}^{J_3\,M_3,J_4\, M_4}=\frac{Sp\left \{
 T_{J_3M_3}(3)T_{J_4M_4}(4)\,F\,
T_{J_1M_1}(1)\,F^+\right\} }{SpFF^+},
\end{equation}
where  $ T_{J_iM_i}(i)$ is the tensor operator of the {\it i}-th particle,
 normalized as \cite{Varschalovich}
\begin{equation}
\label{normtt}
SpT_{JM}^+\,T_{J'M'}=\delta_{JJ'}\delta_{M'M}.
\end{equation}
Using Eqs.(\ref{tfi}), (\ref{f}) and properties of the $T_{JM}$ operators 
\cite{Varschalovich},
we find the following formula
\begin{eqnarray}
\label{ktttgen}
 K_{J_1M_1}^{J_3M_3,J_4M_4}\,Sp FF^+=
\frac{1}{4\pi}\,\sqrt{(2J_1+1)(2J_3+1)(2J_4+1)}\nonumber \\
\times \sum_{S\,S'\,J\,J'\,L\,L'\, J_0\,J_0'}
(2J+1)(2J'+1)\sqrt{(2L+1)\,(2L'+1)\,}\times\nonumber \\  
\times \sqrt{(2S+1)\,(2S'+1)\,(2J_0+1)}
(-1)^{(3(j_1+j_2) - S+L}\times  \nonumber \\
\times (J_0\, -M_1 \, J_1\, M_1 |J_0'\, 0)
(L'\, 0 \, L\, 0\,|J_0' 0)(J_3M_3J_4M_4|J_0-M_1)
\times \ \ \  \nonumber \\
 \left \{\begin{array}{ccc}
j_1 & j_2 & S' \\
S & J_1 & j_1  
\end{array} \right \}
\left \{\begin{array}{ccc}
J' & j_3 & j_4  \\
J & j_3 & j_4  \\
J_0 & J_3 & J_4  
\end{array} \right \}
\left \{\begin{array}{ccc}
J' & S' & L' \\
J & S & L \\
J_0 & J_1 &  J_0' 
\end{array} \right \} a_J^{L\,S} (a_{J'}^{L'\,S'})^* .\ \ \ \ \ \  
\end{eqnarray}
  Due to
 presence of the Clebsch-Gordan coefficient $(L'0L0|J_0'0)$ in 
 Eq. (\ref{ktttgen}) and P-parity conservation
 only even $J_0'$ contribute to the right side of
 Eq.(\ref{ktttgen})
 \begin{eqnarray}
\label{evensum}
L+L'+ J_0' & {\text \,\,\, is\,\,\,  even,\,\,\,\,} &  
 J_0' {\text \,\,\ is \,\,\, even.}
\end{eqnarray}
We are interesting here in the case of $M_4=0$
 and $J_4$  being even, because for this case  the tensor polarizations
 $t_{J_40}$     can be measured via angular momentum distributions
 in the strong two-body decay of the $\Theta^+$ (i.e. {\it 4}-th particle)
 into the nucleon  and pseudoscalar meson \cite{bermanjacob}: $\Theta^+\to 
  N+K$.
 As it follows from Eq.(\ref{ktttgen}), 
 the coefficient $K_{J_1M_1}^{J_3M_3,J_4M_4}$
 is non-zero only under the  following
 condition between  the magnetic quantum numbers
\begin{eqnarray}
\label{tttselection}
  M_1+M_3+M_4=0.
\end{eqnarray}  
 If $J_1+J_3+J_4$
 is {\it even},   the following relations are  valid  for $M_4=0$: 
\begin{eqnarray}
\label{tttsymmetry}
K_{J_1\,+1}^{J_3 \,-1,J_4\,0}=K_{J_1\,-1}^{J_3 \,+1,J_4\,0},\,\, \,
K_{J_1+1}^{J_3\,+1,J_4\,0}=K_{J_1\,-1}^{J_3\,-1,J_4\,0}=0.
\end{eqnarray}
For $J_1=J_3=1$ Eq. (\ref{tttsymmetry})
can be rewritten in terms of Cartesian components  as  
\begin{eqnarray}
\label{tttxyz}
K_{1\,x}^{1\,x,J_40}=K_{1\,y}^{1\,y,J_40}=-K_{1\,-1}^{1\,+1,J_40}=
-K_{1\,+1}^{1\,-1,J_40},\nonumber \\ 
K_{1\,\alpha}^{1\,\beta,J_40}=0 &
 {\text { if $\alpha\not = \beta \,\,\, (\alpha,\beta =x,y,z$). \ \ \ \ \ \ }}
\end{eqnarray}
The coefficient $K_{J_1\,0}^{J_3\,0,J_4\,0}$ can be non-zero only
for $J_1+J_3+J_4$ being { even},
\begin{eqnarray}
\label{ttt000even} 
K_{J_10}^{J_30,J_40}\not =0, 
 & {\text {    if $J_1+J_3+J_4$ is even.}}
 \end{eqnarray} 

   If $J_1+J_3+J_4$ is {\it odd}, then one can find
\begin{eqnarray}
\label{tttsymmetry2}
K_{J_1\, +1}^{J_3\, -1,J_4\, 0}=-K_{J_1\, -1}^{J_3 \, +1, J_4\, 0};\,\,\,
K_{1\,+1}^{1\, -1,J_4 \,0}=
iK_{1\,x}^{1\,y, J_4\,0}=-iK_{1\,y}^{1\, x, J_4\,0},\\ 
K_{1\,x}^{1\,x,J_4\,0}=K_{1\,y}^{1\,y,J_4\,0}=0,\\
\label{ttt000} 
K_{J_1\,0}^{J_3\,0,J_4\,0}=0. 
\end{eqnarray}

 Eq.(\ref{ktttgen}) is rather general and can be used for calculation of 
 the above considered spin observables. So,
 if one put in  Eq.(\ref{ktttgen}) $J_4=0$ and $M_4=0$, then the 
 spin transfer coefficient $K_\alpha^\beta$  ($\alpha,\,\beta=x,y,z$),
 given by  Eq.(\ref{kijgen}), can be obtained from Eq.(\ref{ktttgen}) as 
\begin{equation}
\label{kijfromttt}
K_\alpha^\beta=\frac{1}{3}\sqrt{j_1^{-1}\,j_3^{-1}(j_1+1)(2j_1+1)(j_3+1)
(2j_3+1)(2j_4+1)}K_{1\, \alpha}^{1\,\beta,00}.
\end{equation}
 When substituting $J_1=0$ and $M_1=0$ into Eq.(\ref{ktttgen}), one can
 find spin-tensor  correlation in the final state. In particular, Eq.
(\ref{clkfinal}) follows from Eq.(\ref{ktttgen}) at
 $J_1=0,\, M_1=0,\, J_4=1$.

{  For the spin-singlet initial state $S=S'=0$} the 
 coefficients
$K_{J_1M_1}^{J_3M_3,J_4M_4}$ are non-zero
only for
 unpolarized beam (or target), i.e. $J_1=M_1=0$,
 as it follows from presence of the second
 9j-symbol in Eq. (\ref{ktttgen})
\begin{eqnarray}
\label{kgensingl} 
 K_{K_1M_1}^{J_3M_3,J_4M_4}=\delta_{J_1,0}\,\delta_{M_1,0}
 \,K_{00}^{J_3M_3,J_4M_4},
 & {\text {    if $S=S'=0$.}}
 \end{eqnarray}

 For  $J_1=M_1=0$ one has $J_0=J_0'$, where $J_0'$ is even, according to
Eq. (\ref{evensum}).
 Furthermore, at $j_3=\frac{1}{2}$  
 two values are allowed for the rank $J_3$:
 $J_3=0$ and $J_3=1$. 
 (i) If $J_3=M_3=0$,
 then only even ranks $J_4$ are allowed, as follows from the
 Clebsch-Gordan coefficient $(J_3\,M_3\,J_4\,M_4|J_0\, -M_1)$
 in Eq.(\ref{ktttgen}).
 (ii) For  $J_3=1$ the non-zero 
 coefficients $K_{00}^{1M_3,J_4M_4}$ are allowed for $M_3=-M_4\not  =0$ both
 for even and odd $J_4$. However,
 for the practically interesting case of $M_4=0$ one has $M_3=0$
 due to Eq.(\ref{tttselection}).
  Therefore
  for   $J_1=0$ , $M_4=0$ and {\it even} $J_4$
  the value $J_3=1$ is not allowed due to
 Eq.(\ref{ttt000even}).
 Thus, for
 unpolarized beam and $M_4=0$
 only the coefficients $K_{00}^{00,J_40}$ with even rank $J_4$  are non-zero:
\begin{eqnarray}
\label{kgenstr} 
 K_{00}^{00,J_40}\not =0, 
 & {\text {    if $J_4$ is even.}}
 \end{eqnarray}
 This result is valid { both for the 
 spin-singlet and spin-triplet} initial states.

 \subsubsection{ Tensor polarization for  
 $ j_1+j_2\to j_3+{\vec j_4}$}
 
 Here we consider the tensor polarization of the {\it 4-th} particle,
$t_{J_40}$, when all others particles are unpolarized in the binary reaction
 $1+2 \to 3+{\vec 4}$.
  The tensor polarization of the {\it 4-th} particle can be 
calculated as
\begin{eqnarray}
\label{tJ40}
t_{J_4M_4}=\frac {SpFF^+T_{J_4M_4}(4)}{SpFF^+}.
\end{eqnarray}
Using Eq.(\ref{ttt}) and the  normalization of $T_{JM}$ operator given by 
Eq.(\ref{normtt}), one can find the relation
\begin{eqnarray}
\label{tJ40k}
t_{J_40}=\sqrt{(2j_1+1)(2j_3+1)}K_{00}^{00,J_40}.
\end{eqnarray} 

  For the {\it spin-singlet state $S=S^\prime= 0$},
 the tensor polarization of the 
{\it 4-th}-particle
 does not depend on  the dynamics
 of the reaction $1+2\to 3+4$ at the threshold and
 intrinsic parity of particles and takes the following values
\begin{eqnarray}
\label{tj0singl}
t_{00}=\frac{1}{2},\,\,
 t_{20}=-\frac{1}{2}, & {\text{\,\,\,  for $j_4=\frac{3}{2}$ and $\pi=\pm
    1$}},\\
t_{00}=\frac{1}{\sqrt{6}}, \,\,\,
t_{20}=-\frac{2}{\sqrt{21}}, t_{40}=\frac{1}{\sqrt{7}},
 & {\text{\,\,\,  for $j_4=\frac{5}{2}$
 and $\pi=\pm    1$}},\\
\end{eqnarray}

 For the {\it spin-triplet state $S=1$}
 we find the
 following formulae for $t_{J_40}$ with the even rank  $J_4$:
\begin{eqnarray}
\label{tj0tripl32}
4\pi SpFF^+\,t_{20}=\frac{3}{4}|a_1^1|^2-2|a_2^3|^2-\frac{7}{4}|a_2^1|^2- \\
 \nonumber
-Re\left ( \frac{3}{\sqrt{2}}a_1^1{a_2^3}^*+\frac{3\sqrt{3}}{2}a_1^1{a_2^1}^*-
\frac{\sqrt{6}}{2}a_2^3\,{a_2^1}^*\right ),\,\,\,\,\,\,\,\,\,
\end{eqnarray}
for $j_4=\frac{3}{2}$, $T=1$, $\pi=-1$,
\begin{eqnarray}
\label{tj0tripl32p}
4\pi SpFF^+\,t_{20}=-\frac{5}{4}|a_2^2|^2-\frac{3}{4}|a_1^2|^2- \\ \nonumber
-\frac{1}{2}Re\left ( {\sqrt{15}}a_2^2{a_1^2}^*+\sqrt{30}
a_2^2{a_1^0}^*-{3}\sqrt{2}a_1^2\,{a_1^0}^*\right ), \,\,\,\,\,\,\,\,\,
\end{eqnarray}
for $j_4=\frac{3}{2}$, $T=0$, $\pi=+1$,
\begin{eqnarray}
\label{tj0tripl52m}
4\pi SpFF^+\,t_{20}=-\frac{\sqrt{21}}{3}|a_2^2|^2-\frac{8}{\sqrt{21}}|a_2^4|^2-
\frac{\sqrt{21}}{2}|a_3^3|^2- \\ \nonumber
-Re\left ( \frac{2}{3}a_3^3{a_2^4}^*+\sqrt{2}
a_3^3{a_2^2}^*- 
2\frac{\sqrt{14}}{7}a_2^4\,{a_2^2}^*\right ), \,\,\,\,\,\,\,\,\,
\end{eqnarray}
for $j_4=\frac{5}{2}$, $T=1$, $\pi=-1$,
\begin{eqnarray}
\label{tj0tripl52p}
4\pi  SpFF^+\,t_{20}=
-\frac{5}{\sqrt{21}}|a_2^2|^2-\frac{25}{2\sqrt{21}}|a_3^4|^2-
\frac{12}{\sqrt{21}}|a_3^2|^2-\\ \nonumber
-Re\left ( \sqrt{\frac{10}{7}}a_2^2{a_3^4}^*+2\sqrt{\frac{10}{21}}
a_2^2{a_3^2}^*-
\frac{2}{\sqrt{7}}a_3^4\,{a_3^2}^*\right ), \,\,\,\,\,\,\,\,\,
\end{eqnarray}
for $j_4=\frac{5}{2}$, $T=0$, $\pi=+1$,
\begin{eqnarray}
\label{t40tripl52m}
4\pi SpFF^+\,t_{40}=\frac{\sqrt{7}}{6}|a_3^3|^2+\frac{5\sqrt{7}}{21}|a_2^3|^2+
\\ \nonumber
+Re\left ( \sqrt{\frac{10}{3}}a_3^3{a_2^1}^*+{\frac{10}{3}}
a_3^3{a_2^3}^*-
10\sqrt{\frac{2}{21}}a_2^3\,{a_2^1}^*\right ), \,\,\,\,\,\,\,\,\,
\end{eqnarray}
for $j_4=\frac{5}{2}$, $T=1$, $\pi=-1$,
and 
\begin{eqnarray}
\label{t40tripl52p}
4\pi SpFF^+\,t_{40}=\frac{\sqrt{7}}{42}\bigl \{ -20|a_2^2|^2+27|a_3^4|^2+
22|a_3^2|^2+ \nonumber \\
+10Re \left (\sqrt{30}\,a_2^2{a_3^4}^*+2\sqrt{10}
a_2^2{a_2^3}^*-
2\sqrt{3}a_3^4\,{a_3^2}^* \right ) \bigr \}, \,\,\,\,\,\,\,\,\,
\end{eqnarray}
for $j_4=\frac{5}{2}$, $T=0$, $\pi=+1$. The formulae for $SpFF^+$ are
 given by Eq. (\ref{spurff}).

 The tensor polarization $t_{J0}$ of the odd rank  $J$ vanishes,
 as can be
 seen from Eq. (\ref{kgenstr}).

\subsubsection{Induced tensor polarization for
 ${\vec j_1}+j_2\to j_3+{\vec j_4}$}

 As follows from Eq. (\ref{kgensingl}), for the {\it spin-singlet state}
  the tensor polarization of the
 {\it 4-th} particle  cannot be induced by polarized beam ($J_1\not=
  0$).
 
 For {\it the spin-triplet}
 state $S=S'=1$ and $J_1\not= 0$ 
the non-zero coefficient $K_{J_1M_1}^{00,J_40}\not =0$
 is allowed only for $M_1=0$  as it follows from Eq.(\ref{tttselection}),
 i.e. for longitudinally polarized beam. Furthermore, according to Eq.
 (\ref{ttt000even}), only odd rank $J_4$ is allowed 
for $K_{10}^{00,J_4\,0}$.  
 Since the case with odd $J_4$ is not interesting for our purpose, we do not
 consider here this option in detail. 

\subsection{ Spin-tensor correlation for
$ j_1+j_2\to {\vec  j_3}+{\vec j_4}$}

  Assuming 
  the spin of the {\it 3}-rd particle is 
 $j_3=\frac{1}{2}$, we find for $M_4=0$  
 that the spin-tensor correlation in the final state
 of the reaction $j_1+j_2\to {\vec j_3}+{\vec j_4}$
 for unpolarized beam ($J_1=M_1=0$) and target $(J_2=M_2=0$)
  is non-zero ($K_{00}^{1\,0,\,J_4\,0}\not= 0$)   only for odd rank  $J_4$
(see Eq.(\ref{ttt000even})).

  For $j_3\ge \frac{3}{2}$ and $j_4\ge \frac{3}{2}$
 the spin-tensor correlation coefficient $K_{00}^{J_3\,0,J_4\,0}$ could be 
 non-zero for $J_3$ and $J_4$ being even, however, the  number of independent
  spin amplitudes increases in this case  that makes the relation between
  this coefficient and P-parity non-transparent.

\subsubsection{ The observables 
 for ${\vec j_1}+j_2\to {\vec  j_3}+{\vec j_4}$}} 

 We consider below  the final spin-tensor correlation,
 $K_{1y}^{1y,J0}$,
 induced  by the transversally and longitudinally  polarized  beam
 (or target)
in the reaction $\vec {\frac{1}{2}}+\frac{1}{2}\to \vec {\frac{1}{2}}+
{\vec j_4}$. This coefficient is { non-zero only  for the spin-triplet 
initial state}.

{\it For the even rank $J$} one can find from Eq. (\ref{ktttgen})
\begin{eqnarray}
\label{ttt32m}
4\pi SpFF^+\,K_{1\,y}^{1\,y,20}= \frac{3}{4}(|a_2^3|^2-|a_2^1|^2)+\nonumber 
\\ +Re\bigl
 (\frac{3\sqrt{2}}{8}a_1^1{a_2^3}^* +\frac{\sqrt{6}}{8}a_2^3{a_2^1}^*-
\frac{\sqrt{3}}{4}a_1^1{a_2^1}^*\bigl ) ,\\
4\pi SpFF^+\,K_{1\,z}^{1\,z,20}= -\frac{1}{4}\bigl\{3|a_1^1|^2+2|a_2^3|^2
+3|a_2^1|^2
 +2\sqrt{6}Re\, a_2^3\,{a_2^1}^*\bigl\} ,
\end{eqnarray}
 for $j_4=\frac{3}{2}$, $T=1$, $\pi=-1$,
\begin{eqnarray}
\label{ttt32p}
4\pi SpFF^+\,K_{1\,y}^{1\,y,20}=
 \frac{1}{4}\bigl \{ |a_1^0|^2-|a_1^2|^2+\nonumber 
\\ +Re\bigl
 (\frac{\sqrt{30}}{2}a_2^2{a_1^0}^* -\frac{\sqrt{2}}{2}a_1^2{a_1^0}^*-
\sqrt{15}a_2^2{a_1^2}^*\bigl )\bigl\} ,\\
4\pi  SpFF^+\,K_{1\,z}^{1\,z,20}= -\frac{1}{4}\bigl \{ 2|a_1^0|^2+|a_1^2|^2+
5|a_2^2|^2+ 2\sqrt{2}Re\, a_1^2{a_1^0}^* \bigl\}
\end{eqnarray}
 for $j_4=\frac{3}{2}$, $T=0$, $\pi=+1$,
\begin{eqnarray}
\label{ttt52m}
 4\pi SpFF^+\,K_{1\,y}^{1\,y,20}= 
\frac{2}{\sqrt{21}}\bigl (|a_2^1|^2-|a_2^3|^2\bigr )+\nonumber 
\\ +Re\bigl
 (\frac{2\sqrt{2}}{3}a_3^3{a_2^1}^* -\frac{2\sqrt{3}}{3}\,a_3^3{a_2^3}^*-
\frac{\sqrt{14}}{21}\,a_2^1{a_2^3}^*\bigl ) ,\\
4\pi SpFF^+\,K_{1\,z}^{1\,z,20}= 
-\frac{1}{252}\bigl \{ \sqrt{21}\bigl (49|a_3^3|^2+
 4|a_2^3|^2+ 6|a_2^1|^2\bigr )+ \nonumber 
\\ 
+Re\bigl ( 210\sqrt{2}\,a_3^3{a_2^1}^* +12\sqrt{14}\,a_2^3{a_2^1}^*+
 140\sqrt{3}\,a_3^3{a_2^3}^*\bigl ) \bigl \},
\end{eqnarray}
 for $j_4=\frac{5}{2}$, $T=1$, $\pi=-1$,
\begin{eqnarray}
\label{ttt52p}
4\pi SpFF^+\,K_{1\,y}^{1\,y,20}= \frac{2\sqrt{21}}{13}    
 \bigl (|a_3^4|^2-|a_3^2|^2\bigr )-\nonumber 
\\ -\frac{\sqrt{7}}{42}Re\bigl
 (4\sqrt{10}a_2^2{a_3^4}^* -2\sqrt{30}a_2^2{a_3^2}^*+
 4a_3^4{a_3^2}^*\bigr ) ,\\
4\pi SpFF^+\,K_{1\,z}^{1\,z,20}= -\frac{\sqrt{7}}{252} \bigl \{   
  \sqrt{3}\bigl (21|a_3^4|^2+28|a_3^2|^2+10|a_2^2|^2\bigr )+\nonumber 
\\ + Re\bigl (
30\sqrt{10}\,a_2^2{a_3^4}^* +20\sqrt{30}\,a_2^2{a_3^2}^*+
84\,a_3^4{a_3^2}^*\bigr ) \bigr \},
\end{eqnarray}
 for $j_4=\frac{5}{2}$, $T=0$, $\pi=+1$,
\begin{eqnarray}
\label{ttt52m40}
4\pi SpFF^+\,K_{1\,y}^{1\,y,40}=\frac{\sqrt{7}}{7}\bigl ( 
 |a_2^3|^2-|a_2^1|^2 \bigr ) +\nonumber 
\\ +Re\bigl
 (a_3^3{a_2^3}^* -\frac{\sqrt{6}}{3}a_3^3{a_2^1}^*+
\frac{1}{\sqrt{42}}a_2^3{a_2^1}^*\bigr ) \bigr \},\\
4\pi SpFF^+\,K_{1\,z}^{1\,z,40}=\frac{1}{84}\bigl \{\sqrt{7}\bigl ( 
 16|a_2^3|^2+24|a_2^1|^2 + 49|a_3^3|^2\bigr ) +\nonumber 
\\ + Re \bigl
 (-28 a_3^3{a_2^3}^* -14{\sqrt{6}}a_3^3{a_2^1}^*+
16{\sqrt{42}}a_2^3{a_2^1}^*\bigr ) \bigr \}
\end{eqnarray}
 for $j_4=\frac{5}{2}$, $T=1$, $\pi=-1$,
 and 
\begin{eqnarray}
\label{ttt52p40}
4\pi SpFF^+\,K_{1\,y}^{1\,y,40}=\frac{2}{\sqrt{7}} 
 \bigl (|a_3^2|^2-|a_3^4|^2\bigr ) +\nonumber 
\\ +Re\bigl
 ( -\frac{\sqrt{30}}{21}a_2^2{a_3^4}^* +\frac{\sqrt{10}}{14} a_2^2{a_3^2}^*-
 \frac{\sqrt{3}}{21}a_3^4{a_3^2}^*\bigr ) \bigr \},\\
4\pi SpFF^+\,K_{1\,z}^{1\,z,40}=\frac{\sqrt{7}}{84}\bigl \{ 
 28|a_3^2|^2+ 21|a_3^4|^2+40|a_2^2|^2 -\nonumber 
\\ -Re\bigl
 ( 2\sqrt{30}\,a_2^2{a_3^4}^* +4\sqrt{10}\, a_2^2{a_3^2}^*-
 28\,\sqrt{3}a_3^4{a_3^2}^*\bigr ) \bigr \},
\end{eqnarray}
 for $j_4=\frac{5}{2}$, $T=0$, $\pi=+1$.
%

{\it For the  odd rank J,} 
 one has $K_{00}^{00,J0}=0$, therefore
 the tensor polarization of the {\it 4th} particle 
 $t_{10}$, $t_{30}, \dots$,
 is equal to zero  if the   beam is 
 unpolarized and the polarization of the
 {\it 3-rd} particle is not measured. However, the tensor polarization
 of the odd rank can be induced by 
 polarized beam  in the transitions from the spin-triplet initial state. 
 So, for  $S=1$ we  obtained  the
 following results: 
 \begin{eqnarray}
\label{t1112pp}
K_{1\,x}^{1\, y,10} = -\frac{1}{2}\frac{Im\,a_1^2{a_1^0}^*}
{|a_1^2|^2+|a_1^0|^2},
\end{eqnarray}
for $j_4=\frac{1}{2}$, $T=0$, $\pi=+1$;
 \begin{eqnarray}
\label{t1012pm}
K_{1\,x}^{1\, y,10} = -\frac{\sqrt{3}}{2}
\frac{Im\,a_1^1{a_0^1}^*}
{3|a_1^1|^2+|a_0^1|^2},  
\end{eqnarray}
for $j_4=\frac{1}{2}$, $T=1$, $\pi=-1$;

 \begin{eqnarray}
\label{t1032p}
 4\pi SpFF^+\,K_{1\,x}^{1\, y,10} = -\frac{1}{40} Im\bigl (
 10\sqrt{3}a_2^2{a_1^2}^* +
 5\sqrt{6}\,a_1^0{a_2^2}^* + 3\sqrt{10}a_1^0{a_1^2}^*\bigr ),
\end{eqnarray}
for $j_4=\frac{3}{2}$, $T=0$, $\pi=+1$;
 \begin{eqnarray}
\label{t1032m}
4\pi SpFF^+\,K_{1\,x}^{1\, y,10} = \frac{\sqrt{5}}{40}Im\bigl (
3\sqrt{2}\,a_1^1{a_2^3}^*+  
2\sqrt{3}\,a_2^1{a_1^1}^*+5\sqrt{6}a_2^1{a_2^3}^* \bigr ),
\end{eqnarray}
for $j_4=\frac{3}{2}$, $T=1$, $\pi=-1$;
\begin{eqnarray}
\label{t3052m}
4\pi SpFF^+\,K_{1\,x}^{1\, y,30} =
 \frac{\sqrt{5}}{45}Im\bigl (6\sqrt{7} a_3^3{a_2^3}^*+  
2\sqrt{42}\,a_2^1{a_3^3}^*+ 5\sqrt{6}a_2^1{a_2^3}^* \bigr ),\nonumber \\
4\pi SpFF^+\,K_{1\,x}^{1\, y,10} = -\frac{\sqrt{105}}{210} Im\bigl (
\sqrt{42}\,a_3^3{a_2^3}^*+  
\frac{14}{\sqrt{7}}\,a_2^1{a_3^3}^* +5 a_2^1{a_2^3}^*\bigr ),
\end{eqnarray}
for $j_4=\frac{5}{2}$, $T=1$, $\pi=-1$;
 \begin{eqnarray}
\label{t3052p}
4\pi SpFF^+\,K_{1\,x}^{1\, y,30} = -\frac{1}{3} Im\bigl (
\frac{2\sqrt{6}}{3}\,a_2^2{a_3^4}^*+  
\sqrt{2}\,a_3^2{a_2^2}^* +\frac{14}{\sqrt{15}}a_3^2{a_3^4}^*\bigr ),
\nonumber \\
4\pi SpFF^+\,K_{1\,x}^{1\, y,10} = \frac{\sqrt{7}}{210}Im\bigl (
10\sqrt{3}\,a_2^2{a_3^4}^*+  
15\,a_3^2{a_2^2}^* +7\sqrt{30} a_3^2{a_3^4}^*\bigr ),
\end{eqnarray}
for $j_4=\frac{5}{2}$, $T=0$, $\pi=+1$.

\section{ Full spin structure for the reaction $\frac{1}{2}+\frac{1}{2}\to
\frac{1}{2}+\frac{1}{2}$}
Here  we give the full spin structure 
 of the binary  reaction  for $j_1=j_2=j_3=j_4= \frac{1}{2}$.

\subsection{ The reaction $pn\to \Lambda^0\Theta^+$}

\subsubsection{ The negative parity}
 For the case $T=0$ and $\pi=-1$, one has  $S=0$. 
In this case the amplitude (\ref{tfi}) describes the transition
$^1P_1\to ~^3S_1$ and  can be written as
\begin{equation}
\label{tfipn0}
 M_{\mu_1\,\mu_2}^{\mu_3\,\mu_4}=
 \sum_{\alpha=x,y,z} \,(\chi_{\mu_3}^+\sigma_\alpha\frac{i\sigma_y}{\sqrt{2}}\,
\chi_{\mu_4}^{(T)+})\,(\chi_{\mu_1}^{(T)}\frac{-i\sigma_y}{\sqrt{2}}\,
\chi_{\mu_2}) {\hat k}_\alpha \sqrt{\frac{3}{4\,\pi}}\, a_1^{1\,0}.
\end{equation}
When deriving Eq. (\ref{tfipn0}) from Eq. (\ref{tfi}) we used
 for the Clebsch-Gordan coefficients   
 the formulae given  above after Eq. (\ref{polarsec}).
  The unpolarized cross section corresponding
 to the amplitude (\ref{tfipn0}) takes the 
 following form: 
\begin{equation}
\label{pn0unpol}
 d\sigma_0=\frac{1}{4}\Phi\,\sum_{\mu_1\,\mu_2\,\mu_3\,\mu_4}
 |M_{\mu_1\,\mu_2}^{\mu_3\,\mu_4}|^2=
 \frac{3}{16\pi}\Phi\,|a_1^{1\,0}|^2
 \end{equation}
 that is in agreement with Eq. (\ref{unpols}).
 In order to calculate the polarized cross section we use the 
 density matrix for the spin-$\frac{1}{2}$ particle being
 in the pure spin state $\chi_{\mu_i}$ in the following form:
\begin{equation}
\label{density}
\chi_{\mu_i}\,\chi_{\mu_i}^+ =\frac{1}{2}(1+{\bfg \sigma}\cdot {\bf
  p}_i).
\end{equation}
 Using Eqs.(\ref{density}) and (\ref{tfipn0}) one can write the
  cross section with polarized  both initial and final particles as
\begin{eqnarray}
\label{fullpn0}
d\sigma({\bf p}_1, {\bf p}_2;{\bf p}_3, {\bf p}_4) =
\Phi\, |M_{\mu_1\,\mu_2}^{\mu_3\,\mu_4}|^2=\nonumber \\
=\frac{1}{4} d\sigma_0\, (1-{\bf p}_1\cdot {\bf p}_2)\,[ 1+ {\bf p}_3
\cdot {\bf p}_4-
 2({\bf p}_3\cdot {\hat {\bf k}})({\bf p}_4\cdot {\hat {\bf k}})].  
\end{eqnarray}
 The polarization vectors of the final particles ${\bf p}_3$ 
 and ${\bf p}_4$ are determined by the reaction amplitude
 (\ref{tfipn0}) and can be found using the standard methods
 \cite{bilenky,ohlsen}. After performing  this step and substituting
 the obtained vectors ${\bf p}_3$  and ${\bf p}_4$ 
 into Eq. (\ref{fullpn0}), one can find   the polarized cross section
 $d\sigma({\bf p}_1, {\bf p}_2)$ given by  Eq. (\ref{s0sec}).
 However, the calculation of ${\bf p}_3$ and  ${\bf p}_4$ 
 is not necessarily  and Eq. (\ref{fullpn0}) is 
 sufficient to find all spin
 observables for the reaction described by the amplitude (\ref{tfipn0}).  
 In particular, one can see from Eq. (\ref{fullpn0}) that there is no
  polarization transfer in this reaction ($K_i^j=0,\,
 i,j=x,y,z$), but   there are  spin-spin correlations in both  
 the initial and  final  states. 

\subsubsection{The positive parity}

 For  $T=0$ and  $\pi=+1$  we have  $S=1$. In this case 
  the amplitude in Eq.(\ref{tfi}) describes the transition
$^3S_1-~^3D_1\to ~^3S_1$ and can be
 written as
\begin{equation}
\label{tfipn1}
M_{\mu_1\,\mu_2}^{\mu_3\,\mu_4}
= \sum_{\alpha=x,y,z} \,(\chi_{\mu_3}^+\sigma_\alpha\frac{i\sigma_y}
{\sqrt{2}}\,
\chi_{\mu_4}^{(T)+})\,(\chi_{\mu_1}^{(T)}\frac{-i\sigma_y}{\sqrt{2}}\,
\Pi_\alpha \chi_{\mu_2}),
\end{equation}
where $\Pi_\alpha$ is the following spin operator
\begin{equation}
\label{palpha}
\Pi_\alpha= G\sigma_\alpha +F\,
 {\hat {k}}_\alpha (\bfg \sigma \cdot {\hat {\bf k}})
\end{equation}
with
\begin{equation}
\label{Gpos}
G=\frac{1}{\sqrt{4\pi}}\,(a_1^0+\frac{1}{\sqrt{2}}\, a_1^2)
\end{equation}
and
\begin{equation}
\label{Fpos}
F=-\frac{3}{\sqrt{8\pi}}\, a_1^2.
\end{equation}
The cross section with polarized initial and final particles
 is the following
\begin{eqnarray}
\label{spsp}
d\sigma({\bf p}_1, {\bf p}_2;{\bf p}_3, {\bf p}_4) =
\Phi\, |M_{\mu_1\,\mu_2}^{\mu_3\,\mu_4}|^2= \nonumber \\
\times \sum_{\alpha\, \beta=x,y,z}
 \frac{1}{8}Sp\{\sigma_\alpha\,(1-{\bfg \sigma} \cdot {\bf p}_4)\,
\sigma_\beta (1+{\bfg \sigma} \cdot {\bf p}_3)\}\times\nonumber \\ 
\times \frac{1}{8}Sp\{\Pi^+_\alpha\,(1+{\bfg \sigma} \cdot {\bf p}_2)\,
\Pi_\beta (1-{\bfg \sigma} \cdot {\bf p}_1) \}.\ \ \ 
\end{eqnarray}
Calculating the traces in Eq. (\ref{spsp}), one can find finally
\begin{eqnarray}
\label{spsp2}
d\sigma({\bf p}_1, {\bf p}_2; {\bf p}_3, {\bf p}_4) =
\frac{1}{4}d\sigma_0 \biggl \{ 1+ \bigl(|F+G|^2+2|G|^2\bigr )^{-1}
\biggl \{|F+G|^2({\bf p}_1\cdot {\bf p}_2+{\bf p}_3\cdot {\bf p}_4)-
\nonumber \\
+2\bigl (|F|^2+2 Re \,FG^*\bigr )
\bigl [\,({\bf p}_1\cdot {\hat {\bf k}})({\bf p}_2\cdot {\hat {\bf
    k}}) + ({\bf p}_3\cdot {\hat {\bf k}})({\bf p}_4\cdot {\hat {\bf k}})
\bigr ]+\nonumber \\
+2\bigl (|G|^2+2Re \,FG^* \bigr )({\bf p}_1+ {\bf p}_2)\cdot ({\bf p}_3
+ {\bf p}_4)
-2Re\, FG^* ({\bf p}_1\cdot {\hat {\bf k}} +{\bf p}_2\cdot {\hat {\bf
    k}}) ({\bf p}_3\cdot {\hat {\bf k}}+{\bf p}_4\cdot {\hat {\bf k}})+
\nonumber \\
+2Im\, FG^*\, 
\bigl [   \bigr \{ \bigl (
(\bf p}_3\cdot {\hat {\bf k}})\, [{\bf p}_4\times {\hat {\bf
      k}}]+
({\bf p}_4\cdot {\hat {\bf k}) [{\bf p}_3\times {\hat {\bf k}}]\
 \bigr )
\cdot ({\bf p}_1+{\bf p}_2)\bigr \}
-
\bigl \{\dots  1\leftrightarrow 3, 2\leftrightarrow 4 \dots\bigr \}
 \bigr ]+
\nonumber \\
+4|F|^2({\bf p}_1\cdot {\hat {\bf k}})({\bf p}_2\cdot {\hat {\bf
    k}})
 ({\bf p}_3\cdot {\hat {\bf k}})({\bf p}_4\cdot {\hat {\bf k}})-\nonumber \\
-2\bigl (|F|^2+2 Re\,FG^*\bigr )
\bigl [({\bf p}_1\cdot {\bf p}_2)
 ({\bf p}_3\cdot {\hat {\bf k}})({\bf p}_4\cdot {\hat {\bf k}})+
({\bf p}_1\cdot {\hat {\bf k}})({\bf p}_2\cdot {\hat {\bf
    k}})
 ({\bf p}_3\cdot {\bf p}_4)\bigr ]+\nonumber \\
+2\,Re\, FG^*\bigl [\bigl \{
({\bf p}_1\cdot {\hat {\bf k}})
 ({\bf p}_3\cdot {\hat {\bf k}})({\bf p}_4\cdot {\bf p}_2)+
({\bf p}_1\cdot {\hat {\bf k}})({\bf p}_4\cdot {\hat {\bf k}})
({\bf p}_2\cdot  {\bf p}_3)\bigl \}+ 
\bigl \{ \dots 1\leftrightarrow 2 \dots \bigr \}
 \bigr ]+\nonumber \\
+\bigl(|F+G|^2-2|G|^2\bigr )({\bf p}_1\cdot {\bf p}_2)({\bf p}_3\cdot {\bf
  p}_4)
+2|G|^2\bigl [({\bf p}_1\cdot {\bf p}_3)({\bf p}_2\cdot {\bf
  p}_4)+({\bf p}_2\cdot {\bf p}_3)({\bf p}_1\cdot {\bf
  p}_4)\bigr ]
\biggr \}\biggr \}. \nonumber \\ 
\end{eqnarray}

 The unpolarized cross section for this case is the following
\begin{equation}
\label{pn1unpol}
 d\sigma_0= \frac{1}{4}\Phi\, \left \{
 |G+F|^2+2|G|^2 \right \} = \frac {\Phi}{16\pi}
 3(|a_1^0|^2+|a_1^2|^2),\ \ \ \ \ \ \ \ \
\end{equation}
 were we used  Eqs. (\ref{Gpos}) and (\ref{Fpos}).
 All spin observables for this reaction are contained in Eq. (\ref{spsp2}).
 For example,  the spin-spin correlation coefficient
 $C_{i,j}$ is a factor in front of the product of the polarization vectors
 of the {\it 1}-st and {\it 2}-nd particles, $p_{1\,i}\,p_{2\,j}$,
 in Eq.(\ref{spsp2}). One can find these coefficients as  
\begin{eqnarray}
\label{cij12fg}
C_{x,x}=\frac{|G+F|^2}{|G+F|^2+2|G|^2},\,\,\, 
C_{z,z}=\frac{|G-F|^2-2|F|^2}{|G+F|^2 +2|G|^2},
\end{eqnarray}
 which coincide  with  Eqs. (\ref{cyys1}) and (\ref{czzs1}), respectively.
 The spin transfer coefficient $K_i^j$  
 is the  factor in front of the product of the polarization vectors
 of the {\it 1}-st and {\it 3}-rd  particles, $p_{1\,i}\,p_{3\,j}$,
 in Eq.(\ref{spsp2}). One can find from  Eq.(\ref{spsp2})
\begin{eqnarray}
\label{ckj12fg}
K_x^x=2\frac{|G|^2+ Re\, FG^*}{|G+F|^2+2|G|^2},\,\,\, 
K_z^z=\frac{2|G|^2}{|G+F|^2+2|G|^2},
\end{eqnarray}
 which coincide with  Eqs. (\ref{onehalfplusx})
 and (\ref{onehalfplusz}), respectively.
 One can find also from  Eq. (\ref{spsp2}) 
 the coefficient $K_{1x}^{1y,\,1z}$ as \footnote{Here $K_{1x}^{1y,10}$ is
 defined by Eq. (\ref{ttt}) and the additional factor $2\sqrt{2}$ in 
 Eq.(\ref{kxyz})  follows from
 the relation between the spin-tensor $T_{1M}$ and $\sigma$-matrix:
 $\sigma_M=\sqrt{2}T_{1M}.$} 
\begin{eqnarray}
\label{kxyz}
K_{1x}^{1y,\,1z}=K_{1x}^{1z,\,1y}=-K_{1y}^{1z,\,1x}=
K_{1y}^{1x,\,1z}=\nonumber \\
=\frac{1}{2\sqrt{2}}\frac{2Im\,FG^*}{|G+F|^2+2|G|^2}=
-\frac{1}{2}\frac{Im\,a_1^2{a_1^0}^*}{|a_1^0|^2+|a_1^2|^2},
\end{eqnarray}
that coincides with Eq. (\ref{t1112pp}).
One can find from Eq. (\ref{spsp2}) also  the coefficients
in front of the products $p_{1\,i}\,p_{2\,j}\,p_{3\,k}$, which we denote here
as ${\bar K}_{i,j}^k$. These coefficients 
describe the spin-spin correlation in the initial state, which induces
 the polarization  of the final particle:
\begin{eqnarray}
\label{kzytoxpn}
{\bar K}_{z,y}^x=-{\bar K}_{z,x}^y={\bar K}_{y,z}^x=-{\bar K}_{x,z}^y=
\frac{2Im\,FG^*}{|G+F|^2+2|G|^2}.
\end{eqnarray}
 For the spin-singlet initial state (see Eq.(\ref{fullpn0}))
 these coefficients equal zero.
 
\subsection{ The reaction $pp\to \Sigma^+\Theta^+$}

 In the reaction $pp\to \Sigma^+\Theta^+$ the initial state is an 
 isotriplet, $T=1$.
 Therefore, for $\pi=+1$ one has  $S=0$ whereas for $\pi=-1$ one obtains
 $S=1$.

\subsubsection{The positive  parity}
 
 In  case of $\pi=+1$ 
 the amplitude (\ref{tfi}) describes the transition
 $^1S_0\to ~^1S_0$ and can be written as
\begin{equation}
\label{tfipp0}
 M_{\mu_1\,\mu_2}^{\mu_3\,\mu_4}=
\sqrt{\frac{1}{4\,\pi}}\,(\chi_{\mu_3}^+\,\frac{i\sigma_y}{\sqrt{2}}\,
\chi_{\mu_4}^{(T)+})\,(\chi_{\mu_1}^{(T)}\frac{-i\sigma_y}{\sqrt{2}}\,
\chi_{\mu_2}) {\hat k}_\alpha\,  a_0^{0\,0}.
\end{equation}
The cross section with polarized initial and final particles
 is the following
\begin{eqnarray}
\label{fullpp0}
d\sigma({\bf p}_1, {\bf p}_2;{\bf p}_3, {\bf p}_4) =
\Phi\, |M_{\mu_1\,\mu_2}^{\mu_3\,\mu_4}|^2=\nonumber \\
=\frac{1}{4} d\sigma_0\, (1-{\bf p}_1\cdot {\bf p}_2)\,
(1- {\bf p}_3
\cdot {\bf p}_4).  
\end{eqnarray}
 One can see from this formula that $C_{x,x}=C_{y,y}=C_{z,z}=-1$
 and $C_{x,x}^f=C_{y,y}^f=C_{z,z}^f=-1$, where $C_{i,j}^f$ is the spin-spin
 correlation parameter in the final state.

\subsubsection{The negative parity}

 For $\pi=-1$ one has  $S=1$ in the reaction $pp\to \Sigma^+\Theta^+$.
 In this case 
 the amplitude (\ref{tfi}) can be written as the following 
sum of the two terms $f_0$
 and  $f_1$, describing the transitions 
 $^3P_0\to ~^1S_0$ and  $^3P_1\to ~^3S_1$, respectively,
 \begin{equation}
  M_{\mu_1\,\mu_2}^{\mu_3\,\mu_4}=f_0+f_1,
\end{equation}
 where
\begin{equation}
\label{f0}
f_0=-\frac{1}{\sqrt{4\pi}}\sum_{\beta=x,y,z}\,
(\chi_{\mu_1}^+\sigma_\beta\frac{i\sigma_y}{\sqrt{2}}\,
\chi_{\mu_2}^{(T)+})\,(\chi_{\mu_3}^{(T)}\frac{-i\sigma_y}{\sqrt{2}}\,
\chi_{\mu_4}) {\hat k}_\beta \, a_0^{1},
\end{equation}
\begin{equation}
\label{f1}
f_1=i\,\sqrt{\frac{3}{8\pi}}\sum_{\alpha=x,y,z}\,
\epsilon_{i\,l\,\alpha}\,{\hat k}_i
(\chi_{\mu_1}^+\sigma_l\frac{i\sigma_y}{\sqrt{2}}\,
\chi_{\mu_2}^{(T)+})\
(\chi_{\mu_3}^{(T)}\frac{-i\sigma_y}{\sqrt{2}}\,\sigma_\alpha
\chi_{\mu_4})  \, a_1^{1};
\end{equation}
here $\epsilon_{i\,l\,\alpha}$ is the fully antisymmetric tensor.

\begin{eqnarray}
\label{f0qw}
|f_0|^2=\frac{1}{4\pi}|a_0^1|^2
\sum_{\beta\,, \beta'=x,y,z}
 \frac{1}{8}Sp\{\sigma_\beta\,(1-{\bfg \sigma} \cdot {\bf p}_2)\,
\sigma_{\beta'} (1+{\bfg \sigma} \cdot {\bf p}_1)\}\times\nonumber \\ 
\times \frac{1}{8}Sp\{(1+{\bfg \sigma} \cdot {\bf p}_4)\,
 (1-{\bfg \sigma} \cdot {\bf p}_3) \},\ \ \ 
\end{eqnarray}
\begin{eqnarray}
\label{f1qw}
|f_1|^2=\frac{3}{8\pi}|a_1^1|^2 
\sum_{i,i',l,l',\alpha\,, \alpha'=x,y,z} \epsilon_{i\,l\,\alpha}\,
\epsilon_{i'\,l'\,\alpha'}\, {\hat k}_i\,{\hat  k}_{i'}\times \nonumber \\
\times  \frac{1}{8}Sp\{\sigma_l\,(1-{\bfg \sigma} \cdot {\bf p}_2)\,
\sigma_{l'} (1+{\bfg \sigma} \cdot {\bf p}_1)\}\times\nonumber \\ 
\times \frac{1}{8}Sp\{\sigma_\alpha\,(1+{\bfg \sigma} \cdot {\bf p}_4)\,
\sigma_{\alpha'} (1-{\bfg \sigma} \cdot {\bf p}_3) \},\ \ \ 
\end{eqnarray}

\begin{eqnarray}
\label{f0csf1}
f_0^*\,f_1=-i\frac{1}{4\pi}\,\sqrt{\frac{3}{2}}\, a_1^1\, (a_0^1)^* 
\sum_{i,l,\alpha\,, \beta=x,y,z} \epsilon_{i\,l\,\alpha}\,
{\hat k}_\beta\,{\hat  k}_{i}\times \nonumber \\
\times \frac{1}{8}Sp\{\sigma_l\,(1-{\bfg \sigma} \cdot {\bf p}_2)\,
\sigma_{\beta} (1+{\bfg \sigma} \cdot {\bf p}_1)\}\times\nonumber \\ 
\times \frac{1}{8}Sp\{\sigma_\alpha\,(1+{\bfg \sigma} \cdot {\bf p}_4)\,
 (1-{\bfg \sigma} \cdot {\bf p}_3) \},
\end{eqnarray}
\begin{eqnarray}
\label{f1csf0}
f_0\,f_1^*=i\frac{1}{4\pi}\,\sqrt{\frac{3}{2}}\, a_0^1\, (a_1^1)^* 
\sum_{i,l,\alpha\,, \beta=x,y,z} \epsilon_{i\,l\,\alpha}\,
{\hat k}_\beta\,{\hat  k}_{i}\times \nonumber \\
 \frac{1}{8}Sp\{\sigma_\beta\,(1-{\bfg \sigma} \cdot {\bf p}_2)\,
\sigma_{l} (1+{\bfg \sigma} \cdot {\bf p}_1)\}\times\nonumber \\ 
\times \frac{1}{8}Sp\{\sigma_\alpha\,(1-{\bfg \sigma} \cdot {\bf p}_3)\,
 (1+{\bfg \sigma} \cdot {\bf p}_4) \},
\end{eqnarray}

Performing the traces in Eq. (\ref{spsp}), one can find finally
\begin{eqnarray}
\label{ppspsp2}
d\sigma({\bf p}_1, {\bf p}_2; {\bf p}_3, {\bf p}_4) =
\Phi (|f_0|^2+|f_1|^2 + f_0\,f_1^* + f_0^*\,f_1)= 
\frac{1}{4}d\sigma_0\, \biggl \{1+ \bigl (|a_0^1|^2+3|a_1^1|^2\bigr )^{-1}
\times
\nonumber \\
\times \biggl \{|a_0^1|^2[ ({\bf p}_1\cdot {\bf p}_2)-
({\bf p}_3\cdot {\bf p}_4)]+
(3|a_1^1|^2-2|a_0^1|^2)
({\bf p}_1\cdot {\hat {\bf k}})({\bf p}_2\cdot {\hat {\bf k}})
+3|a_1^1|^2({\bf p}_3\cdot {\hat {\bf k}})({\bf p}_4\cdot {\hat {\bf
    k}})+\nonumber \\
\nonumber \\
+3|a_1^1|^2({\bf p}_1\cdot {\hat {\bf k}}+{\bf p}_2\cdot {\hat {\bf
    k}})
({\bf p}_4\cdot {\hat {\bf k}}+{\bf p}_3\cdot {\hat {\bf k}})+\nonumber \\
+ \sqrt{6} Re \,a_0^1{a_1^1}^* \bigl (
({\bf p}_1\cdot {\hat {\bf k}}+{\bf p}_2\cdot {\hat {\bf k}})
({\bf p}_4\cdot {\hat {\bf k}}-{\bf p}_3\cdot {\hat {\bf k}})
-({\bf p}_1+{\bf p}_2)\cdot ({\bf p}_4-{\bf p}_3)\bigr )
\nonumber \\
%
-\sqrt{6}\, Im \,{ a_0^1\,(a_1^1)^*}
\biggl ( ({\hat {\bf k}}\cdot [{\bf p}_4\times{\bf p}_3])
({\bf p}_1\cdot {\hat {\bf k}}+{\bf p}_2\cdot {\hat {\bf k}})- \nonumber \\
-[{\bf p}_4\times{\bf p}_3]\cdot ({\bf p}_1+{\bf p}_2)
-({\bf p}_2\cdot {\hat {\bf k}})(({\bf p}_4-{\bf p_3})
\cdot[{\hat {\bf k}}\times {\bf p}_1])- \nonumber \\
-({\bf p}_1\cdot {\hat {\bf k}})(({\bf p}_4-{\bf p_3})
\cdot[{\hat {\bf k}}\times {\bf p}_2])
\biggr )- \nonumber \\
-\sqrt{6}\, Re\,{ a_0^1\,(a_1^1)^*}
\biggl ( ([{\bf p}_3\times{\bf p}_4]\cdot[{\hat {\bf k}}\times {\bf p}_1])
({\bf p}_2\cdot {\hat{\bf k}})
+([{\bf p}_3\times{\bf p}_4]\cdot[{\hat {\bf k}}\times {\bf p}_2])
({\bf p}_1\cdot {\hat{\bf k}})\biggr ) -\nonumber \\
-\bigl ( |a_0^1|^2+3|a_1^1|^2\bigr )
({\bf p}_1\cdot {\bf p}_2)({\bf p}_3\cdot {\bf p}_4)-
3|a_1^1|^2\bigl [({\bf p}_1\cdot {\bf p}_2)
({\bf p}_3\cdot {\hat {\bf k}})({\bf p}_4\cdot {\hat {\bf k}})+
\nonumber \\
\bigl ( 2|a_0^1|^2+3|a_1^1|^2\bigr )
({\bf p}_1\cdot {\hat {\bf k}})({\bf p}_2\cdot {\hat {\bf k}})
({\bf p}_3\cdot {\bf p}_4)+\nonumber \\
+3|a_1^1|^2 ({\hat {\bf k}}\cdot [{\bf p}_1\times{\bf p}_3]) 
({\hat {\bf k}}\cdot [{\bf p}_2\times{\bf p}_4])+
({\hat {\bf k}}\cdot [{\bf p}_1\times{\bf p}_4]) 
({\hat {\bf k}}\cdot [{\bf p}_2\times{\bf p}_3])
\bigr ] 
\biggr \} \biggr \}.  \nonumber \\
\end{eqnarray}
 Here the unpolarized cross section is given by
\begin{equation}
\label{unpolmin}
 d\sigma_0=\frac{\Phi}{16\pi}\bigl \{|a_0^1|^2+3|a_1^1|^2\bigr \}.
\end{equation}
 One can see from Eq. (\ref{ppspsp2}) that terms with interference 
 of the amplitudes $a_1^1$ and $a_0^1$  are antisymmetric under
 permutation of the indices 3 and 4. This follows 
 from the fact that the amplitude
 $f_1$ given by Eq. ({\ref{f1}) is symmetric while the amplitude $f_0$
 in Eq. (\ref{f0}) is antisymmetric  under  permutation of the indices
 of the  nonidentical particles  3 and 4. 
 Thus, one can find from the {\it 4}-th line in Eq. (\ref{ppspsp2})
 that  $K_x^x(1,3)=K_x^x(2,3)=
 - K_x^x(1,4)=-K_x^x(2,4)$, where $K_i^j(l,m)$ describes the polarization
 transfer from the {\it l}-th to {\it m}-th particle. 
 All results obtained in the previous sections for the coefficients
 $K_i^j,\, C_{i,j},\, C_{i,j}^f$ and $K_{1\,i}^{1\,j,10}$ $(i,j=x,y,z$)
 for $j_i=\frac{1}{2}$ ($i=1,2,3,4$) and  $\pi=-1$  are contained 
 in Eq.(\ref{ppspsp2}). 
 For example, one can find from Eq. (\ref{ppspsp2}) 
 that the coefficient in front of the product
 $p_{1\,x}\,p_{3\,y}\,p_{4\,z}$, i.e. the
 coefficient $2\sqrt{2}SpFF^+K_{1x}^{1y,10}$,
 coincides with that given by Eq.(\ref{t1012pm}).
 The coefficients in front of the $p_{1\,y}\,p_{3\,y}$ and $p_{1\,z}\,p_{3\,z}$
 coincide with $K_y^y$ and $K_z^z$ in Eqs. (\ref{onehalfminusx}) and
 (\ref{onehalfminusz}), respectively.
 Furthermore, similarly to Eqs. (\ref{kzytoxpn})
 we find from Eq.(\ref{ppspsp2}) the coefficients for
the initial spin-spin correlation with induced polarization of the {\it 3-}rd
particle as follows 
\begin{eqnarray}
\label{kzytoxpp}
{\bar K}_{z,y}^x=-{\bar K}_{z,x}^y={\bar K}_{y,z}^x=-{\bar K}_{x,z}^y=
\frac{\sqrt{6}\,Im\,a_1^1{a_0^1}^*}
{3|a_1^1|^2+|a_0^1|^2}.   
\end{eqnarray} 
   These coefficients are equal to zero for the spin-singlet initial
 state (see Eq. (\ref{fullpp0})).


\section{Discussion and conclusion}

 Our analysis is restricted by the threshold region, assuming
 the s-wave dominance in the relative motion
 of the final particles. This assumption  is expected 
 to be a reasonable approximation up to few MeV of the excess energy
\cite{thomas}.
 To see it one should mention that  for a high momentum transfer,
 what is a specific feature of the reaction
 $NN\to Y\Theta^+$, in the near threshold regime the final partial
 wave with orbital momentum $l$ contributes to the amplitude of the reaction
 as $\sim (p_f/Q)^{l}$, were $p_f$ is the final c.m.s. momentum
 and $Q$ denotes an intrinsic scale of the process which  is determined
 by the  transferred momentum. Therefore, the contribution of the
 higher partial waves $(l\not =0$)  is suppressed at the threshold
 ($p_f< Q$) as compared to $l=0$.
 An additional suppression of the $l\not =0$ partial waves  can be provided
  by centrifugal barrier, if the final state 
 short-range interaction  between the hyperon  $Y$ and $\Theta^+$  is
 strong enough.

 When considering  energy dependence of the spin observables
 near the threshold, one should note that
 if  for a certain spin $S$ (at given $T$ and $\pi$)
 Eq.(\ref{pauli}) holds at the threshold, then above the threshold for
 the same spin $S$ one has 
 the following relation:
\begin{equation}
\label{pauliabove}
(-1)^S=\pi\,(-1)^{T+1}(-1)^l.
\end{equation} 
 Eqs. (\ref{pauli}) and  (\ref{pauliabove}) show
 that only even orbital momenta $l$ in the
 final state are allowed for this spin $S$. In contrast, for the other spin
 ${\bar S}$, which is not allowed at the threshold for the same $T$ and $\pi$
 (for the  $s$ wave in the final
 state), only odd orbital momenta ${\bar l}$ contribute to the final state
 above the threshold. Thus, taking into account
 the above mentioned  $p_f^l$-dependence of the transition
 amplitudes,  one can see that (due to a $p$-wave contribution)
 the  cross section $^{2{\bar S}+1}d\sigma_{\bar M}$
 increases with the excess energy  ($W\sim p_f^2$)
 in the near threshold region   by one power of $W$
   faster  as compared to the 
 $^{2 S+1}\d\sigma_{M}$  cross section.  Here 
 $^{2 S+1}\d\sigma_{M}$ denotes  the  cross section of
 the reaction 
 which is initiated in the NN spin state $|SM>$ \cite{ramachandran1}. 
 For example, for the reaction $pp\to \Sigma^+\Theta^+$
  one has $^3d\sigma_M/p_f \sim const(W)$ and
 $^1d\sigma_{M=0}/p_f\sim W$, if $\pi_\Theta=-1$.
 For $\pi_\Theta=+1$ the energy dependences
 of the singlet and triplet cross sections are interchanged.
 This difference allows one to determine the P-parity
 of the $\Theta^+$ unambiguously
 by measurement of the energy dependence of the
 observables $d\sigma_0\,C_{j,j}$ in the near threshold region 
\cite{hanhart2}, because the unpolarized cross section
 $d\sigma_0$
 can be separated into spin-singlet ($^1\d\sigma_{M=0}$) and spin-triplet
 ($^3d\sigma_M$) cross sections  using
 the  spin-spin correlation parameters
 $C_{x,x},\,C_{y,y}$ and $C_{z,z}$ \cite{ramachandran1}
\footnote{ In particular, the sum 
$\Sigma=C_{x,x}+C_{y,y}+C_{z,z}$ (see Eq.(\ref{summacii}))
is equal to $+1$ ($-3$) for the initial spin-singlet (triplet) state
at  any excess energy $W$ independently on the reaction mechanism.  
 Thus,  a measurement of $\Sigma$  determines the ratio of
 the triplet-singlet difference
 to the unpolarized cross section as
 $\delta=\frac{^3\sigma-^1\sigma}{\sigma_0}=
\frac{1}{2}(1+\Sigma)$. This observable  allows one to determine the P-parity
 of the $\Theta^+$ in limiting cases $\delta=+1$ or $\delta=-1$, which are
 expected to occur near the threshold region.}.
 Similar arguments can be found  for energy dependence of
 the   observables $d\sigma_0\,K_y^y$ \cite{hanhart2},  
 $d\sigma_0\,K_{J_N\,M_N}^{J_Y\,M_Y,\,J_\Theta\,M_\Theta}$
 and $d\sigma_0\,{\bar K}_{x,\, z}^y$,
 although the singlet-to-triplet separation cannot be
 applied here.

 Obviously,   quantitative results
 for spin observables above the threshold can be obtained
 only under additional assumptions about
 the dynamics of this reaction.  
 In order to estimate the upper limit for the excess energy $W$ for P-parity
 determination in the reaction $pp\to \Sigma^+\Theta^+$,
 the authors of   Ref. \cite{hanhart}
 assumed the  (reduced) partial wave amplitudes  
 for different orbital momenta are comparable each with other in
 magnitude.
 Under this assumption they found that
 for excess energy  less than $50$ MeV 
 the sign of the $C_{y,y}$ can be used  
 for unambiguous   determination of P-parity
(for the case of $j_\Theta=\frac{1}{2}$).
 Later on this result was tested
 within a certain  model of the $NN\to Y\Theta^+$ reaction in
 Ref. \cite{hanhart2} and energy dependence
 of $d\sigma_0\,C_{j,j}$ and $d\sigma_0\,K_y^y$
 was  found to be more suitable for P-parity measurement
 within the same region of the excess energy
 $W<50$ MeV.

 Existing model calculations of the $\Theta^+$ production
 in NN-collisions
\cite{liuko,nam,hanhart2} are performed on the basis of
 the kaon exchanges in the Born  approximation, while
 the mechanism of this reaction might be more complicated \cite{karlipkin2}.
 Furthermore,  the initial and final state interactions were neglected in
\cite{liuko,nam,hanhart2}, although these
 can change the
 relative phases of the spin-amplitudes and therefore  provide
 a significant effect of the polarization observables.
 Since at present neither the $\Theta^+$ production mechanism nor
 the  strength of the  final state interaction 
 are  known and the spin of the $\Theta^+$ is not measured,
 it is impossible to construct a quantitatively reliable model for
 the reaction in question. Therefore a general  model-independent
 phenomenological approach for spin-observables 
 at the threshold of the reaction $NN\to Y\Theta^+$ 
 is still an appropriate  method at the first step of analysis, whereas
  energy dependence of the observables in the near threshold region
 can be considered in a largely model independent way according to 
 the method developed in Ref.\cite{hanhart2}.

 The main idea of the work \cite{thomas} for determination of the P-parity of
 the $\Theta^+$ in the ${\vec p}{\vec p}\to \Sigma^+\Theta^+$ reaction is 
 based on the fact that for $\pi_\Theta=+1$ $(-1)$ only spin singlet
 (triplet) initial state is allowed at the threshold 
 and therefore the cross section is non-zero only for antiparallel
 (parallel) spins of the initial protons. This provides very clear 
 signal for $\pi_\Theta$. As it seen
 from Eqs.(\ref{s0sec}) and  (\ref{cyys1}), this signal does not depend on
 the spins of the Y and $\Theta^+$. This conclusion does not depend also on
 the isospin of the $\Theta^+$ when it is equal to $0,1$ and 2.
 Therefore, this reaction seems as a real tool for P-parity
 determination of the $\Theta^+$.
 The problem is, however, connected  with a complexity of such
 kind of experiments.  The cross section near the threshold is suppressed by
 the phase space factor. Furthermore, the requirement of polarized beam 
 and target
 could reduce the luminosity by two orders of magnitude  
 as compared to unpolarized  measurements. Thus, one has to explore  others
 possibilities related to unpolarized or single-polarized experiments.

 In the present analysis of the reaction $NN\to Y\Theta^+$ we use two
 such   opportunities.
 (i) The vector polarization of the hyperon Y and (ii) tensor
 polarization of the $\Theta^+$ (for  $j_\Theta> \frac{1}{2}$) are
 measurable without performing a  secondary scattering.
 So, the vector polarization of the hyperon $Y$ can be measured via its
 weak decay $Y\to N+\pi$, because P-parity violation provides a large
 asymmetry in angular distribution of the decay products.
  The tensor polarization of
 the $\Theta^+$, $t_{J0}$, for even ranks $J$
 can be measured by angular distribution in  the strong decay
 $\Theta^+\to N+ K$.
 For example, according to Ref. \cite{bermanjacob},
 the angular distribution in the helicity frame, $I_{h.f.}(\theta)$,
 determines  the following combinations
 of the spin-density matrix $\rho_{mm'}$ ($m$ and $m'$ are the spin
 projections) of the decaying
 spin-$\frac{3}{2}$ particle:
 $\rho_{\frac{1}{2},\frac{1}{2}}+\rho_{-\frac{1}{2},-\frac{1}{2}}$,
 $\rho_{\frac{3}{2},\frac{3}{2}}+\rho_{-\frac{3}{2},-\frac{3}{2}}$.
 It is easy to find that these combinations  determine the
 $t_{00}$ and  $t_{20}$. A similar conclusion is valid for the spin 
 $\frac{5}{2}$, where the additional combination 
 $\rho_{\frac{5}{2},\frac{5}{2}}+\rho_{-\frac{5}{2},-\frac{5}{2}}$ is appeared
 that allows one to measure the tensor $t_{40}$.
 The angular distribution $I_{h.f.}(\theta)$  in the decay $\Theta^+ \to N+K$
  is determined by the spin of the $\Theta^+$.
 For the spin -$\frac{1}{2}$ decaying particle the  angular distribution 
 $I_{h.f.}(\theta)$ is isotropic \cite{bermanjacob}.
 Thus, measurement of angular distribution
 in the decay $\Theta^+ \to N+K$ allows one to determine  the $\Theta^+$ spin.
 In order to extract the tensor polarization $t_{JM}$
 with $M\not =0$  one needs  to know the longitudinal and transversal
 polarization of the decaying particle \cite{bermanjacob}.
 Such measurements require to perform secondary rescatterings and,
 hence, are  unrealistic for the reaction
 $NN\to Y\Theta^+$ and  not considered here.

  {The main results of this paper related to the threshold
 kinematics of the $NN\to Y\Theta^+$ reaction
  can be summarized as follows.}

{\it Spin-spin correlation in the initial state $C_{i,j}$}.
  Eqs.
 (\ref{s0sec}), (\ref{cyys1})
 and (\ref{cijgen}) 
  allow us to conclude that for $S=1$
 the spin-spin correlation coefficient
  $C_{y,y}$ is always non-negative 
 in the reaction  $1+2\to 3+4$ 
  at the  threshold  independently of the spins $j_3$ and $j_4$.
  On the contrary, for $S=0$, the spin-spin correlation coefficients
  $C_{x,x}=C_{y,y}=C_{z,z}$ are equal to $-1$ for arbitrary spins of
  final particles.
  The obtained  result
  allows one to determine unambiguously the P-parity of the $\Theta^+$
  by measurement of  $C_{y,y}$ 
  in the reaction
  $pp\to \Sigma^+\,\Theta^+$. The   total isospin of this channel is
  fixed ($T=1$) therefore the spin $S$ of the initial nucleons is directly
  related to the  P-parity $\pi_\Theta$ of the $\Theta^+$ as
  $(-1)^S=\pi_\Theta$. In the reaction
  $pn\to \Lambda^0 \Theta^+$ one has  either  $(-1)^S=-\pi_\Theta$,
  if the isospin
  of the $\Theta^+$ is even ($I_\Theta=0$, $2$), or
  $(-1)^S=\pi_\Theta$, if $I_\Theta=1$.
  Therefore, both the P-parity and the isospin of the $\Theta^+$ 
  can be determined unambiguously  by combined measurement of
   $C_{y,y}$  in these two reactions.

{\it Spin-spin correlation in the final state $C_{i,j}^f$}, given by
  Eqs.(\ref{clkfinal}), (\ref{cft1pmin})-(\ref{cf352t0min}),
 can be used  in principle for  P-parity determination in a binary reaction
 $1+2\to {\vec 3}+{\vec 4}$, 
 if the vector  polarizations of final particles are
 measurable in some way. In the reaction 
 $NN\to {\vec Y}{\vec \Theta^+}$ 
 only the polarization of the hyperon $Y$ is self-analyzing
 via its weak decay, but a possibility  to measure  the polarization of the 
 $\Theta^+$ is very questionable.
 Non-zero spin-tensor correlation parameters 
 in the final state  $K_{00}^{10,J_\Theta0}$ are allowed only for the odd rank
 $J_\Theta$ and,
 therefore, also  unlikely can be measured in the $NN\to Y\Theta^+$ reaction.

 {\it The spin-transfer coefficient $K_i^j $. }
  The vector polarization transfer  is strongly
  correlated with the  P-parity of the $\Theta^+$ and the isospin of
  the NN-channel. The coefficient $K_y^y$ is zero for the spin singlet
  and non-zero for the spin-triplet NN-state. In other words,
 in the reaction $pp\to \Sigma^+\Theta^+$ one has  $K_y^y\not =0$ 
 for $\pi_\Theta=-1$
  and $K_y^y =0$ for $\pi_\Theta=+1$. 
  In the reaction $pn\to \Lambda\Theta^+$ 
  the relation between $K_y^y $  and $\pi_\Theta$ is inverted, if the
  $\Theta^+$ is an isoscalar.
 For $j_\Theta=\frac{1}{2}$  we found $K_z^z\ge 0$ independently of
 $\pi_\Theta$. For the isovector $\Theta^+$, the spin
  observables in the reactions $pp\to \Sigma\Theta^+$ and  $pn\to
  \Lambda\Theta^+$ are identical.

{\it Tensor polarization of the $\Theta^+$},
  $t_{J0}$, for even rank $J$ can be measured by analysis of the angular
 distribution in the strong  decay $\Theta\to N+K$ that
 does not require polarized  beam and/or  target.
  At given $j_\Theta >\frac{1}{2}$, as it seen from Eqs. 
 (\ref{tj0tripl32}), the coefficients 
 $K_{00}^{00,J0}\sim t_{J0}$ at $J=0,\,2,\,4, \dots$ 
 are different for $\pi=-1$ and $\pi=+1$. 
  The absolute value and  the sign of
 this difference depend on the dynamics of the reaction
 $NN\to Y\Theta^+$. Hence, it is necessary  to study these
 observables within definite mechanisms of the reaction.  
 At last, the values of $t_{J0}$
 at even rank $J$
 give a definite  restriction to the spin of the $\Theta^+$.

{\it The spin-transfer coefficient  $K_{1y}^{1y,J0}$.}
 If the spin of the $\Theta^+$ is higher than $\frac{1}{2}$,
 then there are additional possibilities for P-parity determination.  
 For even rank $J$ 
 the spin-tensor correlation induced by polarized beam,
 $K_{1y}^{1y,J0}$
  and $K_{1z}^{1z,J0}$ are zero for the spin singlet  and
 non-zero for the spin-triplet
 states. Therefore, similarly to the $K_y^y$ coefficient, 
 for given isospin  and P-parity 
   of the $\Theta^+$, the coefficient 
 $K_{1y}^{1y,J0}$ 
 is non-zero only for one isospin channel of the reaction
 $NN\to Y\Theta^+$ and equals zero for another one.
 This  statement is valid for any spin of the $\Theta^+$. 
 Therefore,  measurements of both  coefficients $K_{1y}^{1y,J0}$
 and $K_y^y$  in
 the reactions $pp\to \Sigma^+\Theta^+$ and $pn\to \Lambda\Theta^+$ 
 give a strong  test for the P-parity determination.

{\it The spin-transfer coefficients ${\bar K}_{x,z}^y$} given by Eqs. 
(\ref{kzytoxpn}) and  (\ref{kzytoxpp}) are non-zero only for the
 spin-triplet initial state. These coefficients are measurable
 in the reaction ${\vec N}{\vec N}\to {\vec  Y}\Theta^+$ and,
 therefore, can
 be used  as an additional test in the P-parity determination.

 In conclusion, the formalism for double and triple spin correlation
 parameters of a binary reaction is derived at the threshold
 for arbitrary spins of the participating particles.
 For the spin-$\frac{1}{2}$ baryons the obtained formulae are checked 
 by different methods.
 The  formalism  can be
 applied for P-parity determination of the $\Theta^+$ in the reaction
 $ NN\to Y\Theta^+$ and any narrow resonance with arbitrary spin.

{\bf Acknowledgments}. The author would like to thank 
 A.~Dorokhov, C.~Hanhart, V.L.~Lyuboschitz,  N.N.~Nikolaev, N.M.~Piskunov
 and O.V.~Teryaev  for  stimulating discussions. Financial support
 from Forschungszentrum J\"ulich and the warm hospitality 
 of Institut f\"ur Kernphysik-II, were a part of this work was carried out, 
 is gratefully acknowledged.    

\section*{Appendix}
\setcounter{equation}{0}
\renewcommand{\theequation}{A.\arabic{equation}}
 Here we derive Eqs. (\ref{kijgen}) and (\ref{k}) for
 the spin-transfer coefficient $K_i^j$.
 From Eqs. (\ref{tfi}), (\ref{f})  and (\ref{kijff}) we find
 \begin{eqnarray}
\label{kijclebsh}
SpF F^+\, K_\lambda^\kappa= \sum_{L\,M_L\, S\,M_S\,J\,M_J\atop
  m_1\,m_2\,m_3\,m_4}
\sum_{L'\,M_L'\, S'\,M_S'\,J'\,M_J'\atop  m_1'\,m_2'\,m_3'\,m_4'}
 a_J^{LS}(a_{J'}^{L'\,S'})^*\, Y_{LM}^*({\bf k})Y_{L'M'}({\bf k})
\times \nonumber \\
(j_1m_1j_2m_2|S\,M_S)(j_3m_3j_4m_4|J\,M_J) (SM_S\,LM_L|J\,M_J)\times
 \nonumber \\
\times (j_1m_1'j_2m_2'|S'\,M_S')(j_3m_3'j_4m_4'|J'\,M_J')
  (S'\,M_S'\,L'\,M_L'|J'\,M_J')\,
 Q_\lambda^\kappa, \ \ \ \ \ \ 
\end{eqnarray}
where $Q_\lambda^\kappa$ is the following trace
  \begin{eqnarray}
   \label{mlk}
 Q_\lambda^\kappa= Sp\biggl \{
 \chi_{j_1\,m_1}^+(1) \chi_{j_2\,m_2}^+(2)
\chi_{j_3\,m_3}(3)\,\chi_{j_4\,m_4}(4) S_\lambda(1)\times  \nonumber \\
\times  \chi_{j_4\,m_4^\prime}^+(4) \chi_{j_3\,m_3^\prime}^+(3)
\chi_{j_2\,m_2^\prime}(2)\,\chi_{j_1\,m_1^\prime}(1) S_\kappa(3)\biggr \}.
\end{eqnarray}
Using the following representation \cite{Varschalovich} for the
 spin operator $S_\lambda(j)$ of  the {\it k}th
  particle ($k=1,\dots, 4$) with the spin $j$
\begin{equation}
\label{spin}
S_\lambda(k)=\sqrt{j(j+1)}\sum_{m \, m'}(j m\, 1\lambda | j\, m')
  \chi_{j\,m^\prime}(k) \chi_{j\,m}^+(k)
\end{equation}
and the  relations
\begin{eqnarray}
\label{tr1}
Sp\left \{ \chi_{j\,m}^+(k) \chi_{j\,m}(k)\right \}=
Sp\left \{ \chi_{j\,m}(k) \chi_{j\,m}^+(k)\right \}=
\delta_{m\,m^\prime},
\end{eqnarray}
one can find
\begin{eqnarray}
\label{tr2}
Sp\left \{  \chi_{j\,m^\prime }^+(k) S_\lambda(k) \chi_{j\,m}(k)\right
\}=
Sp\left \{  \chi_{j\,m}(k)\chi_{j\,m^\prime}^+(k) S_\lambda(k)
 \right \}= \nonumber \\
=(-1)^\lambda\sqrt{j(j+1)}\,(jm\, 1-\lambda|j\,m^\prime).
\end{eqnarray}

Making use  Eqs. (\ref{tr2}) and (\ref{tr1}), one can present Eq. (\ref{mlk})  as
\begin{equation}
\label{mlk2}
Q_\lambda^\kappa=(-1)^{\lambda+\kappa}\, \delta_{m_2\,m_2^\prime}\, \delta_{m_4\,m_4^\prime}
\sqrt{j_1(j_1+1)j_3(j_3+1)}(j_3m_3^\prime\,1\, -\kappa|j_3m_3)
(j_1m_1\,1 -\lambda|j_1\,m_1^\prime).
\end{equation}
 After substituting Eq. (\ref{mlk2}) into Eq. (\ref{kijclebsh}), one has to perform 
 summation over the spin projections in Eq. (\ref{kijclebsh}). It  can be done
 using the following relations
\begin{eqnarray}
\label{sumclebsch}
\sum_{m_1\,m_1'\,m_2}(j_1m_1\,j_2m_2|S\,M_S)
 (j_1m_1'\,j_2 m_2|S'\,M_S')(j_1m_11-\lambda|j_1m_1') = \nonumber \\
 (-1)^{j_1+j_2+S+1}\sqrt{(2j_1+1)(2S+1)} 
(SM_S\,1\,-\lambda|S'\,M_S')
 \left \{ \begin{array}{ccc}
j_1 & j_2 & S \\
S' & 1 & j_1  
\end{array} \right \},\ \ \ \ \ \  \\
%
\sum_{m_3\,m_3'\,m_4}(j_3m_3\,j_4m_4|J\,M_J)(j_3m_3'\,j_3 m_4|J'\,M_J')
  (j_3m_3'1 -\kappa |j_3m_3)=\nonumber \\
=(-1)^{j_3+j_4+J'+1}\sqrt{(2j_3+1)(2J'+1)} (J'M_J'\,1-\kappa|J\,M_J)
\left \{\begin{array}{ccc}
j_3 & j_4 & J' \\
J & 1 & j_3  
\end{array} \right \}, \ \ \ \ \ \ \\
%
\sum_{M_S\,M_S'\,M_J\,M_J'}
(SM_S\,LM_L|JM_J)(S'M_S'\,L'M_L'|J'M_J')(SM_S\,1\, -\lambda|S'M_S')
(J'M_J'\,1 -\kappa|J\, M_J)=\nonumber \\
=\sum_{J_0\,M_0} (-1)^{L-S-S'-M_L'-1}(2J+1)\sqrt{(2S'+1)(2J'+1)}\times \nonumber \\
 \times(L'\,-M'LM|J_0M_0)(1\,-\lambda \, 1-\kappa|J_0\,M_0)
\left \{\begin{array}{ccc}
J & S & L \\
J' & S' & L' \\
1 & 1 &  J_0 
\end{array} \right \}. \ \ \ \ \
\end{eqnarray}
After that Eq. (\ref{kijclebsh}) can be written  as Eq. (\ref{kijgen}).

In order to obtain  Eqs. (\ref{k}), we use here  the relation between the
Cartesian and spherical components of the spin operator \cite{Varschalovich}
\begin{eqnarray}
\label{cartesian}
S_{+1}=-\frac{1}{\sqrt{2}}(S_x+iS_y),\nonumber \\
S_{-1}=\frac{1}{\sqrt{2}}(S_x-iS_y),\nonumber \\
S_0=S_z.
\end{eqnarray}
From Eqs. (\ref{kijff}) and (\ref{cartesian}) one can find
\begin{eqnarray}
\label{axial}
K_{+1}^{-1}=-\frac{1}{2}\left [K_x^x+K_y^y+iK_y^x-iK_x^y\right ],\nonumber \\
K_{-1}^{+1}=-\frac{1}{2}\left [K_x^x+K_y^y-iK_y^x+iK_x^y\right ],\nonumber \\
K_{+1}^{+1}= \frac{1}{2}\left [K_x^x-K_y^y+iK_y^x+iK_x^y\right ],\nonumber \\
K_{-1}^{-1}= \frac{1}{2}\left [K_x^x-K_y^y-iK_y^x-iK_x^y\right ].
\end{eqnarray}

As was noted in the text after  Eq. (\ref{kijgen}), the angular momentum $J_0$ is
 even. Therefore one has   $(1-\lambda 1\lambda|J_00)=(1\lambda
 1-\lambda|J_00)$.
It allows one to find from Eq. (\ref{kijgen})  the following  relation
\begin{equation}
\label{axial2}
K_{-1}^{+1}=K_{+1}^{-1}.
\end{equation}
Using Eqs.(\ref{kijgen}), (\ref{axial}) and (\ref{axial2}) one finds
$K_{+1}^{-1}-K_{-1}^{+1}=i(K_x^y-K_y^x)=0$ and 
$K_{+1}^{+1}-K{_1}^{-1}=i(K_x^y+K_x^y)=0$, therefore
\begin{equation}
\label{kxy}
 K_x^y=K_y^x=0.
\end{equation}
Using relations $K_{-1}^0=K_{+1}^0=K_0^{+1}=K_0^{-1}=0$, one can find
\begin{equation}
K_x^z=K_y^z=K_z^x=K_z^y=0.
\end{equation}
At last, taking into account the relation
$K_{+1}^{+1}=0$, which follows from  Eq. (\ref{kijgen}),
 we find from the third equation in Eqs. (\ref{axial})
\begin{equation}
\label{kxxkyy}
K_x^x=K_y^y.
\end{equation}
From Eqs. (\ref{kxxkyy}), (\ref{kxy}) and (\ref{axial}) one can find 
 Eqs. (\ref{k}).

}
\end{document}